# Nanoscale friction controlled by top layer thickness in [LaMnO$_3$]$_m$/[SrMnO$_3$]$_n$ superlattices

**N.A. Weber\*, M. Lee, F. Schönewald, L. Schüler, V. Moshnyaga, M. Krüger and C.A. Volkert**

Niklas A. Weber\*, Florian Schönewald and Cynthia A. Volkert

*Institute of Materials Physics, University of Göttingen, Friedrich-Hund-Platz 1, Göttingen 37077, Germany*

Miru Lee and Matthias Krüger

*Institute for Theoretical Physics, University of Göttingen, Friedrich-Hund-Platz 1, 37077 Göttingen, Germany*

Leonard Schüler, Vasily Moshnyaga

*1st Physics Institute, University of Göttingen, Friedrich-Hund-Platz 1, 37077 Göttingen, Germany*

**Graphical Abstract**

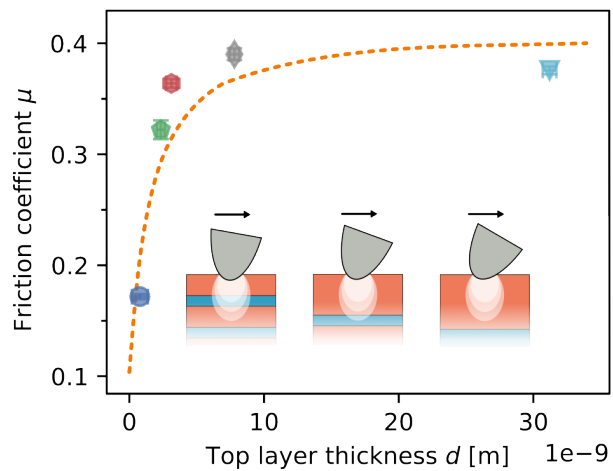

**Abstract**

We conducted lateral force microscopy measurements on seven $[LaMnO_3]_m/[SrMnO_3]_n$ superlattices with varied layer thicknesses. We observe that the friction forces and the friction coefficients initially increase with increasing $LaMnO_3$ top layer thickness, followed by saturation when the top layer thickness exceeds a few nanometers. These observations clearly demonstrate that sliding friction is affected by sub-surface material properties to a depth of several nanometers and is not just determined by dynamics in the contact interface. We argue that the sub-surface dissipated energy is governed by damping in the elastically strained volume below the AFM tip, an effect which we estimate via thermoelasticity. The absence of a correlation between friction and the thermal resistivity of our superlattices shows furthermore that high-frequency phonons and heat conduction do not play a role in determining friction. Our observations thus demonstrate that friction can be tailored by sub-surface material properties.

# I. INTRODUCTION

During the relative motion of two bodies in contact, kinetic and mechanical energy is converted into heat. This energy dissipation process is called friction and understanding its origins has been a long-standing issue in research, technology, and society. One of the main challenges of the last centuries has been that friction could only be studied on macroscopic scales, which kept the underlying dissipation mechanisms hidden for a long time. The emergence of atomic force microscopy (AFM) in the 1980s [1] facilitated the study of friction at the single nano-asperity level [2–14]. In the case of nanoscale low-wear, dry sliding friction, two mechanisms have been predominantly discussed: 1. frictional losses due to phonon excitation by mechanical interactions between the tip and surface, and 2. dissipative interactions linked to the electron system. To unravel the different contributions to sliding friction, several literature studies have so far systematically varied temperature [2–4,13], contact pressure [5] or bias voltages [6–9] to alter the electronic and electrostatic interactions between the tip and the samples.

In our recent experimental studies on manganite thin films [3,6] we found evidence that electronic and electrostatic interactions are not sufficient to account for the observed changes in friction. Instead, it could be shown that phononic and vibrational contributions are the dominant factor contributing to energy dissipation at a nanoscale sliding contact. However, it remained hidden which phonon modes and wavelengths dominate dissipation and to what depth beneath the surface the material properties influence friction. These findings motivated the studies presented here, of nanoscale friction measurements on $[LaMnO_3]_m/[SrMnO_3]_n$ superlattices.

The question of how material properties below the surface affect friction has received considerable attention [10–12,15–23]. Friction measurements on layered materials, such as graphene, $MoS_2$, $NbSe_2$, $WS_2$, $WSe_2$ and h-BN report a decrease in friction with an increase in layer count [11,12,15]. To explain these observations, several mechanisms have been proposed [15–17,19,22] that are related to the so-called puckering effect, which only is expected in materials with weak interlayer forces, and changes in the elastic properties. At the same time, studies [11,15] shows that the measured frictional forces approach a bulk value for increasing layer numbers, but do not specify the parameters that determine the length scale at which saturation occurs. In this regard, a viscoelastic model developed by Lee et al. [19], which attributes top layer thickness dependent friction to viscous dissipation inside the evanescent waves excited in the top layer by a vibrating tip provides a possible explanation. In their study, the authors relate the saturation length to the decay length of such evanescent waves.

In this manuscript, we present AFM friction measurements on seven $[LaMnO_3]_m/[SrMnO_3]_n$ superlattice films, with varying layer thicknesses in the range of $m = 2$ to 80 atomic layers (or 0.8 to 32 nm). Our main finding is an increase in friction with the top layer thickness, which saturates after the layer thickness exceeds a few nanometers. We further observe no correlation between friction and the thermal properties of our films [24,25]. This enables us to rule out thermal conductivity, and thus high-frequency phonons, as the dominant contribution to friction in our systems. From the observed thickness dependence of the friction coefficients, we develop a model connecting friction to the energy losses occurring in the stressed volume near the contact interface through thermoelastic damping. Our model provides a possible explanation for the dependence of friction on the surrounding material properties, and is consistent with the observed linear correlation between friction and normal force. We thus propose that friction can be actively controlled through tailored material selection, such as the thermal expansion coefficient, which consequently opens new possibilities for control of friction.

## II. MATERIALS & METHODS

**Superlattice Films**

Six epitaxial superlattice films of $[LaMnO_3]_m/[SrMnO_3]_n$, where $m$ refers to the number of $LaMnO_3$ unit cell layers, and $n$ to the number of $SrMnO_3$ unit cell layers, were grown using metal organic aerosol deposition technique (MAD) [26]. An overview of the $m$ to $n$ ratio of the films studied here can be found in Table 1. Each superlattice stack was terminated with a layer of $[LaMnO_3]_m$ to ensure equivalent chemical composition on the surface. The layer thicknesses and periodicity were monitored during deposition using in-situ ellipsometry measurements (see Section SI-1.1). The total superlattice film thicknesses were kept constant at around 30 nm. Additionally, a single layer $m = 80$ thick $[LaMnO_3]_{80}$ film was prepared using the same technique. All films were deposited on 5 x 5 x 0.5 mm$^3$ (100)-oriented $SrTiO_3$: 1.0 at. % Nb $K \leq 0.5°$ substrates.

| m [3.9 Å] | n [3.9 Å] | Film thickness [nm] | Thermal resistivity $1/\kappa$ [mK/W] |
|---|---|---|---|
| 2 | 2 | 32 | 0.91(2) |
| 4 | 2 | 34 | 3.27(7) |
| 6 | 6 | 32 | 0.43(2) |
| 8 | 4 | 31 | 1.84(7) |
| 14 | 7 | 32 | 1.23(7) |
| 20 | 10 | 35 | 0.99(7) |
| 80 | 0 | 32 | 0.77(7) |

*Table 1: Layer thicknesses and thermal resistivity values [24] of the seven superlattices studied. Individual thicknesses of layers $m$ and $n$ were obtained from XRR measurements [27]. Thermal resistivity values were approximated from thermal transient reflectivity measurements [24] (see Section SI-1.5). Depending on the $m/n$ ratio, LaMnO$_3$ is either cubic ($m/n = 1$) or a rhombohedral ($m/n = 2$).*

The films were characterized after deposition using standard x-ray methods including $\theta - 2\theta$ x-ray diffraction (XRD), small-angle x-ray scattering (XRR), and x-ray photoelectron spectroscopy (XPS) measurements. Additionally, the magnetic properties of the films were characterized by temperature dependent measurements of the magnetic moment (SQUID).

XRD measurements show Kiessig fringes [28] near the substrate peak, and no reflexes related to impurities (see Section SI-1.2). XRR measurements show the superlattice reflections and have intensities consistent with an interface roughness of $RMS \leq 0.6$ nm for all films (see Section SI-1.3). XPS measurements have been performed on all samples using a Kratos Axis Supra spectrometer to investigate the chemical compositions of the near surface regions (see Section SI-1.4).

Thermal resistivities $1/\kappa$ of the films studied here were estimated by using measurements made previously on films deposited with the same protocols and in the same deposition chamber [24]. Specifically, the thermal resistance $1/\kappa$ of various [LaMnO$_3$]$_m$/[SrMnO$_3$]$_n$ films, including films with the same $m$ and $n$ ratios as those shown here, were obtained using optical transient thermal reflectivity measurements and exhibit a linear dependence on the interface density. The linear relation was used to estimate the thermal resistivity $1/\kappa$ of the films used here and are listed in Table 1 (see Section SI-1.5). The validity of this comparison was additionally supported by SQUID measurements (see Section SI-1.6), that revealed magnetic properties of the films studied here in good agreement to those in earlier work [24,29].

**AFM Measurements**

AFM measurements were performed at room temperature ($T = 293\ K$) under UHV ($p \approx 2 \times 10^{-10}$ mbar) conditions using an Omicron VT-AFM/STM. For lateral force microscopy measurements, standard rectangular, single crystalline silicon cantilevers (*NANOSENSORS™ PPP-LFMR* [30]) with a nominal tip radius $r \leq 10$ nm were used. The cantilevers exhibit normal spring constants $k_N$ ranging from 0.88 to 1.3 Nm$^{-1}$ and lateral spring constants $k_T$ ranging from 132.3 to 177.4 Nm$^{-1}$. The latter were calculated based on geometric properties provided by the manufacturer and material parameters from literature [31]. Normal spring constants were measured using the so called *Sader method* [32,33] which is based on a known cantilever geometry and measuring the quality factor of the oscillation resonance. A representative SEM-image of an AFM cantilever and tip can be found in Section SI-2.

Friction measurements were performed using lateral force microscopy measurements, whereby the torsion of the cantilever is measured as it is dragged over the film surface at a constant applied normal force $F_N$ and velocity ($v = 250$ nm s$^{-1}$). The measured torsion is directly proportional to the lateral forces $F_L$ during sliding. To separate changes in topography from friction effects, so-called friction loops – trace and retrace scanning along the same line on the film surface – are recorded $F_F = 1/2 \cdot (F_{L,trace} - F_{L,retrace})$ [34].

Friction forces $F_F$ were obtained by averaging over 50 friction loops that were recorded within a 100 x 250 nm$^2$ surface region with a point density of 1 nm$^{-1}$ along the fast, and 0.5 nm$^{-1}$ along the slow scan direction for each normal load and film. To assess how robust and reproducible the friction measurements are and to determine the order of magnitude of possible variations, the measurements were repeated several times on randomly selected areas on the film surface. In addition, to minimize wear during the friction experiments, the applied normal forces $F_N$ were kept below 30 nN and any possible changes in the contact were monitored through adhesion measurements.

By measuring the frictional forces on the same surface area as a function of an applied load, a friction coefficient $\mu$ between the tip and the specimen can be determined using a modified Amontons relation,

$$F_F = \mu \cdot F_N + F_{F0}, \text{ (Eq. 1)}$$

where the term $F_{F0}$ is the non-vanishing frictional force at $F_N = 0$ N and is related to adhesion forces $F_A$ between tip and film, that are non-negligible at the nanoscale [34].

Adhesion forces were equated to the pull-off forces obtained by averaging over 50 force-distance curves. Significant changes in adhesion forces before and after probing the frictional properties can indicate changes in tip geometry due to wear, changes in the surface chemistry, or electrostatic forces do to triboelectrification [35]. To ensure that our friction measurements did not alter the surface morphology or affect the friction forces measured, an overview scan of the measurement area was performed after friction measurements.

To measure several films consecutively without the need to change the cantilever or the force calibration, up to four films were adhered to a Omicron stainless steel sample plate using a silver-filled epoxy (*EPO-TEK H21D*) to ensure good thermal and electric conductivity. Additionally, a small droplet of contact silver (*ACHESON 1415 G3692*) was placed at the edge of each film, far away from the area studied, to electrically contact the sample surfaces.

## III. RESULTS & DISCUSSION

**XPS Measurements**

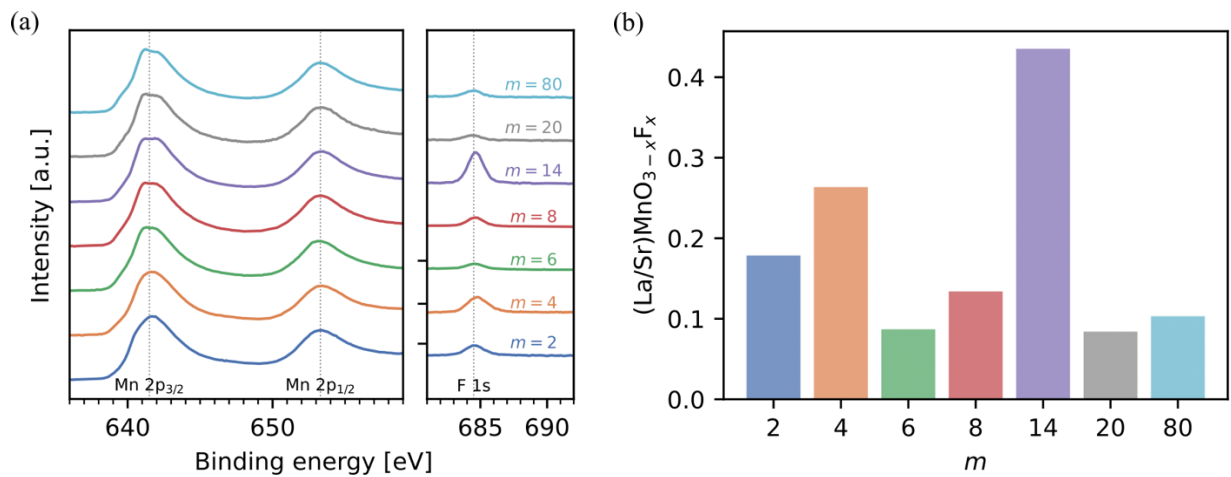

*Figure 1: (a) Excerpts from XPS spectra showing the Mn2p$_{1/2}$ and Mn2p$_{3/2}$ peaks at binding energies of 653 eV and 641 eV, as well as the F1s peak at a binding energy of 685 eV. (b) The fluorine concentration in the near surface region of the superlattice samples is determined from the ratio of the Mn to F-peak areas. The full spectra, as well as further details on the evaluation, are presented in the supporting information (see Section SI-1.4).*

XPS measurements were performed to confirm the chemical composition of the manganite film surfaces. In addition to the La, Sr, and Mn peaks, the spectra's exhibit an additional peak at 684.45 eV which can be assigned to fluorine (F) (see Figure 1(a), full spectrum, see Section SI-1.4).

In manganites, F is known as a source of contamination in sample preparation involving Teflon components [36] or as a dopant in LaMnO$_3$/SrMnO$_3$ [37]. If F is present as a contaminant on the sample surface, an F1s binding energy of 689.8 eV is expected, together with a C1s signal at 292.5 eV, reflecting C-F bonds [36]. However, the measured F 1s binding energy is significantly smaller and no component in the C1s peaks at 292 eV is observed (see Section SI-1.5). Therefore, we can rule out that F is present as a surface contaminant. Instead, the binding energy of the F1s peak matches with SrF$_2$ (684.6 eV) [38] or LaF$_3$ (684.5 eV) [39], suggesting that F dopes the material by replacing oxygen in our films [37].

The fluorine content in our films was estimated from the F 1s and Mn 2p peak intensities using standard methods (see Figure 1(b) and Section SI-1.4). It can be observed that the films with top layer thickness of $m = 4$ and $14$ exhibit a particularly high F concentrations compared to the other five superlattice films.

In systems such as $SrMnO_{3-x}F_x$ fluorine doping can result in ferromagnetic behavior, the occurrence of a charge ordered magnetic transition, local distortions of Mn octahedra [37], and a reduced electrical conductivity [40]. Similarly, F doping can introduce a short-range magnetic order in $LaMnO_{2.8}F_{0.2}$ [41] or can lead to ferroelectric properties of $LaMnO_2F$ [42], thus altering structural, electric, and magnetic properties. However, such a variation is not detect in our SQUID measurements (see Section SI-1.6).

Since we will observe in the subsequent sections that the significantly higher fluorine concentration in samples $m = 4$ and $14$, will have a significant influence on the measured frictional forces, we will discuss them separately.

**Friction Measurements**

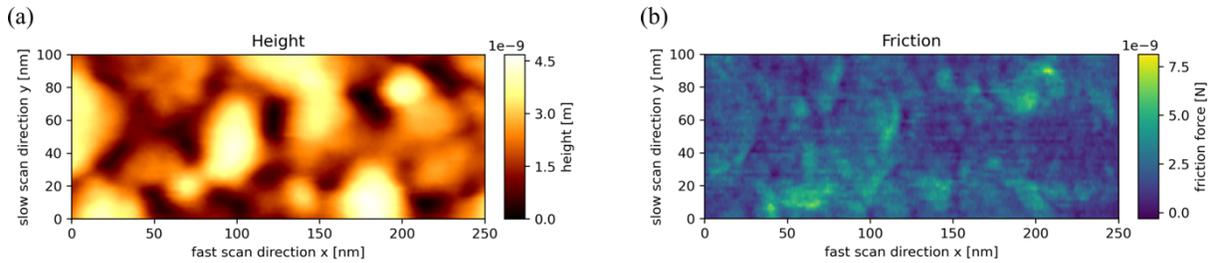

*Figure 2: (a) Topography and (b) corresponding friction maps of a $[LaMnO_3]_2/[SrMnO_3]_2$ surface. The topography map shows a smooth surface with a roughness of $RMS = 0.4$ nm, and no apparent correlation between the calculated friction map and topography.*

Figure 2 depicts a topography map and the corresponding friction map, that is calculated from the lateral traces measured for a $[LaMnO_3]_2/[SrMnO_3]_2$ film. A Spearman correlation coefficient of $r_S = 0.014$ between the height and friction maps and of $r_S = 0.08$ between the friction and height gradient along x-direction, indicating no correlations between friction and topography, and support the validity of the lateral force method [34] to measure friction. Similar surface topographies and friction maps were obtained on all $[LaMnO_3]_m/[SrMnO_3]_n$ films and show comparably low correlations $r_S < 0.15$ and surface roughness $RMS \leq 0.5$ nm. All films show small variations in the friction force ($\pm 2$ nN) on a length scale of about 50 nm (see Figure 2). Shifts in the absolute values of friction forces from map to map are on the order of $\pm 5$ nN and indicate small changes in surface chemistry, variations in top layer thickness (see Table 1), carbon-based surface residues and adsorbed water. Occasional larger shifts in friction forces ($\pm 10$ nN) are sometimes observed when comparing different measurements made with different cantilevers and can attributed to changes in tip shape. Representative friction and topography maps for all films can be found in the supporting information (see Section SI-3).

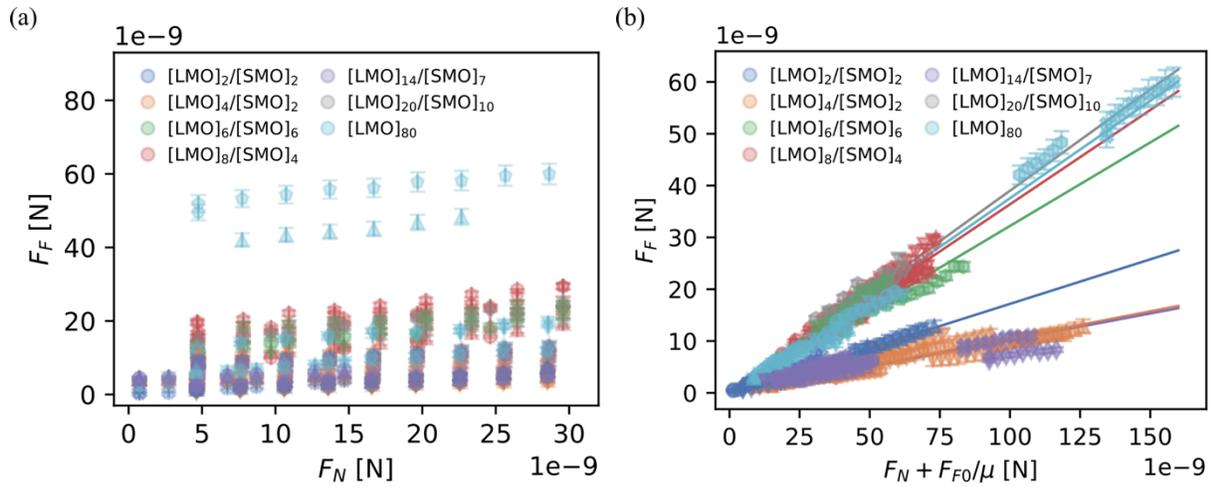

Figure 3: Friction forces $F_F$ measured for each $[LaMnO_3]_m/[SrMnO_3]_n$ film show (a) linear dependencies on the applied normal load $F_N$ with slopes $\mu$ that are fairly constant for each film and intercepts $F_{F0}$ which vary strongly from map to map. (b) The friction forces for each film collapse onto a single proportional line when plotted against $F_N + F_{F0}/\mu$.

Friction forces $F_F$ for each film were measured for applied normal loads $F_N$ ranging from 1 to 30 nN by averaging over a single friction map (250 nm × 100 nm region). The friction increases linearly with normal force, irrespective of the $m/n$ ratio and fluorine content (Figure 3(a) and Section SI-4.1). Linear regressions (Equation 1) to the friction data were performed to obtain the non-zero friction force intercept $F_{F0}$ as well as a friction coefficient $\mu$ (slope) for each individual friction versus normal force measurement (Figure 3(a) and Section SI-4.1). While the slopes $\mu$ did not show a large variation between measurements on a given sample, the values for $F_{F0}$ vary from map to map and show no clear dependence on nominal layer thicknesses or other material properties (see Section SI-4.1). However, when divided by the friction coefficient $F_{F0}/\mu$, they correlate well with the pull-off forces obtained from force-distance-curves (see Section SI-4.2), in good agreement with various models for elastic contacts that include adhesion [43,44]. In fact, we find that the friction forces for each film are reasonably well described as proportional to the sum of the applied normal load $F_N$ plus the adhesive force as estimated by $F_{F0}/\mu$ (Figure 3(b) and Section SI-4.3). Average friction coefficients $\bar{\mu}$ for each film were obtained from the best fit slopes to these data (Figure 3(b)).

The fact that the friction forces for each sample are approximately described by a master curve that is linear in the applied normal load plus $F_{F0}/\mu$, is in good agreement with the Derjaguin-Muller-Toporov (DMT) model for adhesive elastic contacts [44,45]. This model provides a good description of adhesive contacts between two hard materials and includes adhesion forces both inside and outside the contact area [44]. The model assumes that the contact profile remains the same as for a Hertzian contact, although this is clearly an over-simplification, particularly in the limit of small applied forces. Several different descriptions of the adhesive contribution have been considered, but in the simplest approach, the adhesive force is added to the applied force in the Hertz contact equations [43]. In this model, the adhesive force is equal to the pull-off force and given by $F_A = 2\pi R \gamma$ where $R$ is the radius of the AFM tip (assumed spherical) and $\gamma$ is the work of adhesion. The measured pull-off forces (between 3 and 75 nN, Figure SI-4.3) agree well with this model for a tip radius $R = 10$ nm, yielding values of $\gamma$ between 0.05 and 1.20 J m$^{-2}$ [43,46].

The friction coefficients $\mu$ (Figure 3(a)) and $\bar{\mu}$ (Figure 3(b)) obtained from the best fit slopes to the data for the different films are between 0.1 and 0.4, and are comparable to AFM friction coefficients on other manganite thin films [3,6]. The coefficients are plotted against the thermal resistivity $1/\kappa$ and top layer thickness $d$ of each film (Figure 4), where the two films with the highest fluorine content are indicated by red data points and have noticeably lower friction coefficients than most other films. There is no clear trend of the friction coefficients with thermal resistivity (Figure 4(a)), even after considering the two different $m/n$ ratios and possible effects of F doping. In contrast, the plot of the friction coefficients versus the top layer thickness (Figure 4(b)) shows a clear trend, when the highly F doped specimens are excluded: the friction coefficient increases with top layer thickness and then saturates above a thickness of around 5 − 10 nm. This is also seen in Figure 3(b), where the slope of the friction force versus $F_N + F_{F0}/\mu$ systematically increases with top layer thickness.

The lack of a clear dependence of the friction coefficient on the thermal resistivity (Figure 4(a)) suggests that friction is not controlled by the GHz to THz range phonons which determine the thermal conduction in these materials [24]. Instead, it seems likely that friction is dominated by the much lower frequency mechanical vibrations that are present during AFM measurements, such as observed in studies of sonolubricity [47]. These mechanical vibrations range from washboard frequencies on the order of 1 kHz up to cantilever mechanical resonances as high as 100 kHz. Presumably, excess THz to GHz phonons are not strongly generated by the mechanical vibrations, and so do not play a role in determining friction. Furthermore, these results indicate that the dissipation of heat by thermal conduction does not appear to play a decisive role in friction.

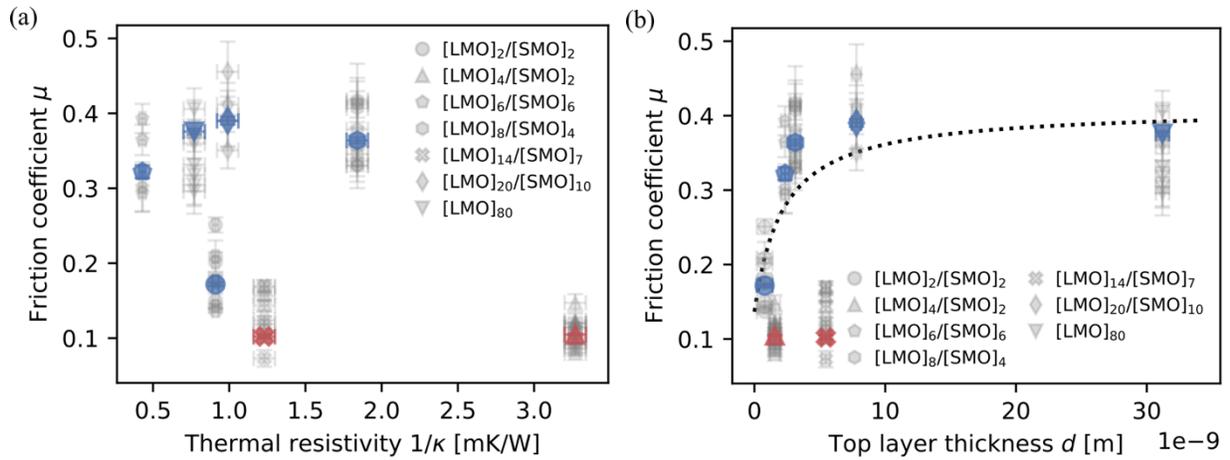

Figure 4: Friction coefficients of the seven different superlattice films plotted against (a) thermal resistivity $\kappa^{-1}$, and (b) LaMnO$_3$ top layer thickness. The friction coefficients $\mu$ obtained from each individual series of measurements are shown as gray data points (from Figure 3(a) and Section SI-4.1); the blue or red data points are the average values $\bar{\mu}$ for each film (from Figure 3(b) and Section SI-4.3). The red data points indicate the films with high fluorine concentrations. There is no apparent trend of friction coefficient with thermal resistivity (a), while the coefficients of the undoped films (blue data points) increase and saturate with LaMnO$_3$ top layer thickness (b). The dashed line is a fit with Equation 3.

Instead, the friction coefficients show a clear trend with LaMnO$_3$ top layer thickness, pointing to the role of the sub-surface material properties on friction, since the surfaces of all film samples are the same. Further, the fact that the friction coefficient saturates for layers thicker than around 10 nm indicates that material properties to this depth affect energy dissipation at the sliding AFM contact. The idea that surrounding material properties affect friction is not new, and several studies support it [2–14]. However, to our knowledge, all these studies involve either a change in electronic properties (and possibly in electrostatic forces) or a phase transformation, so that the bonding states of the atoms at the surface are likely changed. In the results we present here, the surface roughness and chemistry of all samples are the same, yet the material properties at a depth between 0.8 and 10 nm below the surface cause the friction coefficient to change by up to a factor of more than two.

In most discussions of friction, it is assumed that the friction force results from dissipative processes occurring exclusively in the contact area. The ground-breaking model of Bowden and Tabor [48], which was able to provide a physically reasonable interpretation of the seemingly nonphysical Amontons/Coulomb friction laws, is based on the idea that friction is simply proportional to the true area of contact between the sliding materials. For AFM measurements of friction, $F_F \propto \pi a^2$, where $a$ is the contact area radius between the tip and the sample. However, our results show unequivocally that contributions from the surrounding materials also contribute. In addition, for an elastic contact between a tip of radius $R$ and a flat sample, the contact area scales sub-linearly with the normal force, $a^2 \propto F^{2/3}$, which contradicts the measured linear dependence observed here.

We therefore propose that the stressed regions of the material surrounding the contact interface contribute to friction through internal damping, an effect which we estimate from thermoelasticity. According to Hertz elastic contact theory [49,50], the stress fields under a spherical AFM tip fall off over several nanometers for typical AFM tip radii and normal forces, consistent with our observation that the friction "feels" to a depth of around 10 nm. Through thermoelastic coupling, the stress fields produced by a sliding tip would -- under adiabatic conditions – cause local temperature changes $\Delta T = -(\alpha/\rho c_p)T_0 \Delta\sigma$, where $\alpha$ is the linear thermal expansion coefficient, $\rho c_p$ is the volumetric specific heat at constant pressure, $T_0$ is the ambient temperature and $\Delta\sigma$ is the sum of the principle stresses [51]. The resultant temperature gradients produce heat flow which can be as large as $Q = -\varepsilon_{th} V \Delta\sigma$, where $V$ is the stressed volume, and $\varepsilon_{th} = T_0 \alpha$ the thermal strain. For an AFM tip sliding with speed $v_s$, the power $P = F_F \cdot v_s$ of sliding friction that can be dissipated by the thermoelastic effect can be estimated as,

$$P = F_f v_s = 2f \cdot \beta \cdot \varepsilon_{th} \cdot V \cdot \Delta\sigma \approx v_s \cdot \frac{8}{\pi} \cdot \beta \cdot \varepsilon_{th} \cdot (F_N + F_A) \text{ (Eq. 2)}$$

where $f$ is the frequency of loading and unloading under the tip. The factor of 2 after the second equality originates from the assumption that the heat flow is generated both on loading and unloading. Eq. (2) assumes that these processes are entirely irreversible, and Eq. (2) is to be seen as an upper limit of the contributions to friction from thermoelasticity. How closely realistic conditions may come to it will be investigated in future work using microscopic models. For continuous, smooth sliding, $f$ is given by $v_s/(2a)$. The quantity $V\Delta\sigma$ can be approximated as $8a^3 p_m$ where $p_m = (F_N + F_A)/(\pi a^2)$ is the mean contact pressure within the contact [49,50] and falls off over a distance that scales with $a = (3R \cdot (F_N + F_A)/(4E^*))^{1/3}$, which is the radius of the contact area between a tip of radius $R$ and the flat sample. Using the nominal AFM tip radius of $R = 10$ nm, the contact radius will range between 0.52 and 2.83 nm, and the contact pressure will range between 1.18 and 6.56 GPa for our experiments [52–54]. $E^*$ is the so-called indentation modulus depending on elastic properties of both the sample and the tip and is 97 GPa for our system [52,54]. Since the elastic constants of the layer materials differ by only about 7%, we can ignore effects of the composite structure on the stress fields. The estimation of the stressed volume under the tip by $V\Delta\sigma$ is a strongly simplified; the reasons being that the stresses are heterogenous and the exact form of the adhesive forces in the contact area are unknown.

In addition to providing an explanation for the dependence of the friction force on the sub-surface material properties according to Equation 2, the thermoelastic damping due to the moving stress fields under the sliding tip predicts a linear dependence of friction on normal force, with a contribution from thermoelasticity to the friction coefficient that can be as large as $\mu_{TE} = 8/\pi \cdot \varepsilon_{th}$. A necessary requirement is that the diffusion occurs much faster than tip motion, which is the case for our system ($D_{th} > v_s/f$ where $D_{th}$ is the thermal diffusivity and is of order $7 \times 10^{-7}$ m² s⁻¹ for [LaMnO3]$_m$/[SrMnO3]$_n$ superlattices [24]. The thermal expansion coefficients of the superlattices are of order $1.5 \times 10^{-6}$ K$^{-1}$ [55] so that the contribution of thermoelasticity to friction is $\mu \approx 0.012$. Surprisingly this is only a factor of 10 smaller than the measured friction coefficients and comes closer to explaining the experimental values than the available models based on electronic excitations [7,56], which are many orders of magnitude too small. Any structural or electronic defects will increase the damping above what is discussed here. Contributions from dissipative processes in the contact area will also contribute to the friction, although the exact mechanism continues to be a topic of debate [45,57–59].

For the superlattice films investigated here, any contributions to the friction from contact area dissipation will be the same for all films, while the contribution from the thermoelastic effect will depend on the layer thicknesses and the depth of the stress fields, which depend on the normal force. We can write an approximate expression for the thermoelastic contribution to friction,

$$\mu_{TE}(d) = (\mu_{LMO} - \mu_0) \cdot g(d/a) + \mu_0 \qquad \text{Eq. (3)}$$

where $\mu_{LMO} = 8/\pi \cdot \beta \cdot \varepsilon_{LMO}$ and $\mu_0 = 8/\pi \cdot \beta \cdot \varepsilon_0$. The function $g(d/a)$ represents the fraction of the thermoelastically damped volume that lies in the top LaMnO$_3$ layer and changes from 0 to 1 as $d$ increases from 0 to infinity. A plot of the measured friction coefficients versus $d/a$ shows the data can be described by the simple functional form $g(d/a) = d/(d+a)$ (see Section SI-4.4) although there is no apparent physical basis for this dependence. The same functional form also gives a good description of the integrated stress fields under a spherical tip [49,50] (see Section SI-5). Moreover, the functional form is in quantitative agreement with a viscoelastic model developed by Lee et al. [19], which attributes top layer thickness dependent friction to viscous dissipation inside the evanescent waves set up in the top layer by a vibrating tip.

We use Equation 3 to fit the data in Figure 4(b), which shows that the expansion coefficients of LaMnO$_3$ must be larger than that of SrMnO$_3$ by more than a factor of 3 to account for the measurements, since the friction coefficient increases from around $\bar{\mu}$ = 0.17 for the thinnest layers, where the stressed volume under the tip extends over several different layers, up to $\bar{\mu}$ = 0.4, where the stressed volume lies entirely in the LaMnO$_3$ top layer. We have found two literature sources for thermal expansion coefficients in these materials, confirming that the thermal expansion coefficient of LaMnO$_3$ ($\alpha_{LMO} \approx 14.6 \times 10^{-6}$ K$^{-1}$ [55]) is larger than of SrMnO$_3$ ($\alpha_{SMO} \approx 8.75 \times 10^{-6}$ K$^{-1}$ [60]), although not quite by a factor of 3.

As far as we can tell, this publication and a recent theoretical study [19] are the first to discuss contributions of thermoelastic damping to measured friction, although thermoelastic damping in AFM cantilevers is routinely considered when interpreting dynamic AFM measurements [13]. We have found no literature discussing thermoelastic damping in the stressed volume around the contact area, which yields an estimate for vibrational damping [61,62]. Although the thermoelastic contributions are somewhat too small to fully account for the measured friction forces, they do provide a simple and clear framework for explaining why friction forces often scale linearly with normal forces and why they are dependent on sub-surface material properties. Internal damping has indeed been suggested theoretically as a main contribution to damping of a weakly coupled probe [19]. In this study the found dependence on layer thickness is strikingly similar to what is observed in our experiments and to what is estimated from thermoelasticity. Ref. [19] identifies the local dissipation as originating from evanescent phonons, which can be interpreted as being excited by a moving stress field. How far these mechanisms share the same physical origins poses an exciting question for future work.

Macroscale contacts are composed of many individual elastic and plastic contacts between micrometer- to nanometer-scale asperities, each of which is expected to behave similarly to an AFM contact. As such, thermoelastic damping should play an important role in macroscale friction as well. The stress fields under a macroscale contact reach to a depth comparable to the individual asperity contact area dimensions, on the order of nanometers or micrometers, so that the contributions of thermoelastic damping relative to dissipation in the contact area should be the same in macroscale contacts as in the AFM studies. Of course, friction at macroscale contacts also involves contributions from plasticity, so the separation of all the contributions and prediction of friction coefficients becomes even more complex.

## IV. CONCLUSION

We report AFM-based friction measurements on seven [LaMnO$_3$]$_m$/[SrMnO$_3$]$_n$ superlattice films, finding an increase in friction with increasing in LaMnO$_3$ top layer thickness for chemically identical surfaces. When the layer thickness exceeds several nanometers, friction saturates to the value found for bulk LaMnO$_3$. This observation shows a connection between friction and material properties up to several nanometers below the surface, and thus contradicts the widely accepted picture that friction at an elastic contact only occurs by dissipative processes in the contact area [45,57–59].

We explain the contribution of the surrounding material to friction by thermoelastic damping in the stressed material near the contact interface. Using simple scaling arguments from elastic contact theory, we can show that thermoelastic damping provides a non-negligible contribution to measured friction forces at both nanoscale and macroscale contacts. Furthermore, thermoelastic damping provides a possible explanation for the observed linear dependence between AFM friction and normal forces, which is otherwise expected to show sublinear behavior if only dissipative processes in the contact area are considered.

The work presented here introduces thermoelastic damping as an important contribution to AFM friction. In contrast to the dissipative processes occurring directly in the contact, which are highly sensitive to real-time conditions, the contributions from thermoelastic and internal damping can be tailored through material selection, thus revealing possible strategies to reliably control friction.

## V. ACKNOWLEDGEMENTS

This work was funded by the German Research Foundation (DFG) 217133147/SFB 1073, Project A01 and Project A02. The authors thank D.R. Baer M.H. Engelhard from Pacific Northwest National Laboratory for discussion and help in interpreting the fluorine contamination observed in XPS measurements. We would like to thank R.L.C. Vink for critical discussions of our measurement results. We thank T. Brede for x-ray characterization of all films and SEM measurements of cantilevers. Additionally, we thank A. Wodtke and F. Güthoff from the Max Planck Institute for Multidisciplinary Sciences Göttingen for providing access and to maintenance of the Omicron VT-AFM system.

# Supporting Information

*N.A. Weber\*, M. Lee, F. Schönewald, L. Schüler, V. Moshnyaga, M. Krüger and C.A. Volkert*

**Niklas A. Weber\*, Florian Schönewald and Cynthia A. Volkert**

*Institute of Materials Physics, University of Göttingen, Friedrich-Hund-Platz 1, Göttingen 37077, Germany*

**Miru Lee and Matthias Krüger**

*Institute for Theoretical Physics, University of Göttingen, Friedrich-Hund-Platz 1, 37077 Göttingen, Germany*

**Leonard Schüler, Vasily Moshnyaga**

*1st Physics Institute, University of Göttingen, Friedrich-Hund-Platz 1, 37077 Göttingen, Germany*

# SI-1 Material Characterization

## SI-1.1 Ellipsometry Measurements

The phase shift angle Δ as a function of time $t$ is monitored in situ during metal organic aerosol deposition technique (MAD). Using the known deposition rate $v_{LMO} = 0.37$ u.c./s and $v_{SMO} = 0.25$ u.c./s from previous studies, a superlattice with the desired $m$ to $n$ was deposited. The layer thicknesses were additionally characterized by XRR measurements (see Section 1.4).

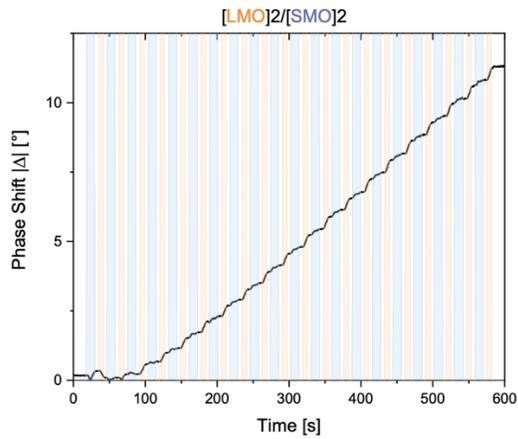
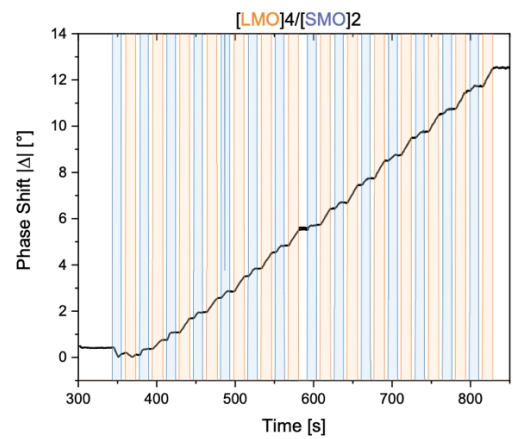
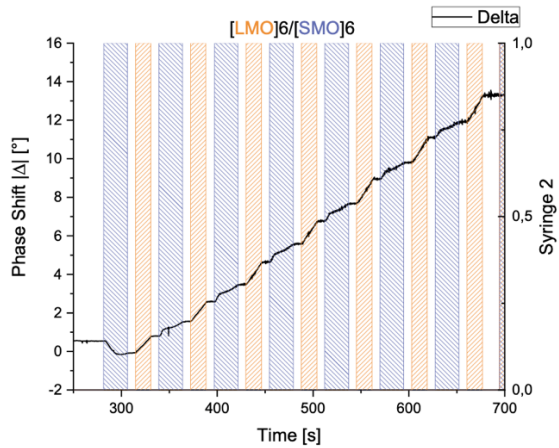
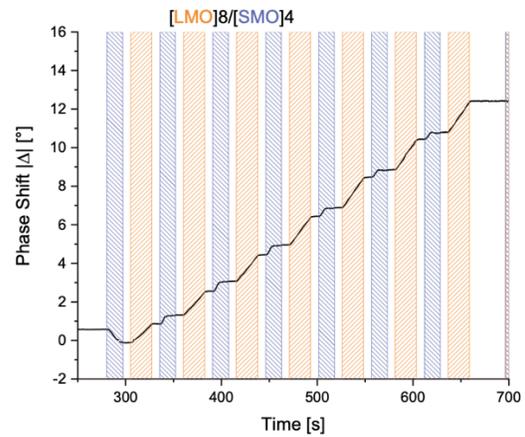

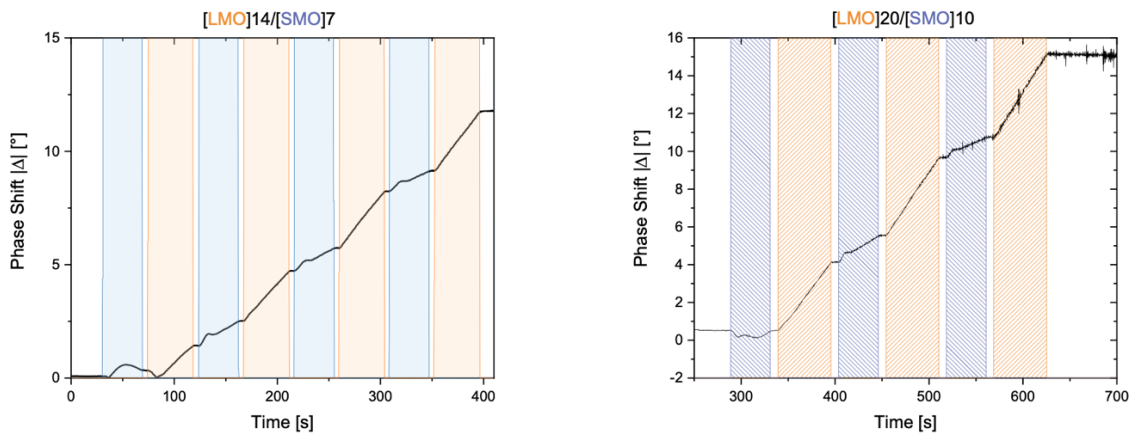

*Figure S1: Ellipsometry phase shift angles Δ measured during deposition of the superlattice films. The orange shaded regions correspond to LaMnO deposition and blue to SrMnO deposition.*

### SI-1.2 XRD Measurements

Θ-2θ x-ray diffraction measurements were performed using a Bruker D8 Discover equipped with a Cu $K_\alpha$ source. The measurements confirm the superlattice periodicity through superstructure reflections labeled by the numbers in below (see Figure S2). Additionally, peaks that correspond to the characteristic wavelength of tungsten (W) can be identified, which are due to aging of the copper source. The peak positions indicated by dashed lines in Figure S2 are listed in Table S1.

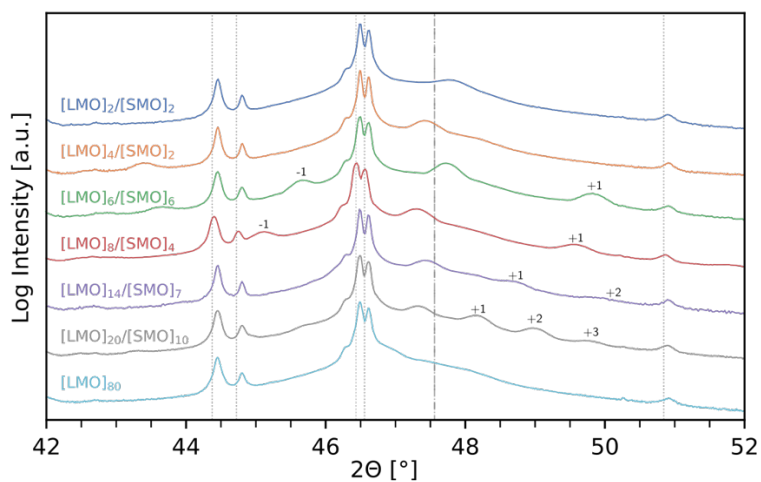

*Figure S2: $\Theta - 2\theta$ x-ray diffraction pattern from the films studied showing peaks corresponding to the substrate SrTiNbO as well as to the superlattice structure. The peak positions indicated by dashed lines are listed in Table S1.*

| Material | (hkl) | Wavelength | 2Θ |
|---|---|---|---|
| SrTiNbO | (002) | Cu Kα$_1$ | 46.433 |
| | | Cu Kα$_2$ | 46.555161 |
| | | W Kα$_1$ | 44.3734354 |
| | | W Kα$_2$ | 44.7219802 |
| | | W L | 50.843 |
| SrMnO | (200) | Cu Kα$_1$ | 47.559 |
| LaMnO | (002) | Cu Kα$_1$ | 46.9399526 |
| | | W Kα$_1$ | 43.4585829 |

*Table S1: Peaks shown in the $\Theta - 2\Theta$ diffraction pattern can be assigned to the SrTiNbO substrate and the LaMnO/SrMnO layers of the superlattice films. We also identify diffraction peaks that correspond to the characteristic wavelength of tungsten (W).*

SI-1.3 XRR Measurements

X-ray reflection measurements were carried out using a Bruker D8 Advance equipped with a Cu K$_\alpha$ source. To determine layer thicknesses and periodicities as well as interface roughness, the data was modeled using the GenX software [S1]. In this software the samples were modeled as a stack of alternating LaMnO$_3$ layers with thickness $d_{LMO}$ and roughness $\sigma_{LMO}$, and SrMnO$_3$ layers with thickness d$_{SrMnO}$ and roughness $\sigma_{SMO}$. The measured data with the corresponding fits are depicted in Figure S3. The parameters obtained from fitting are listed in Table S1.

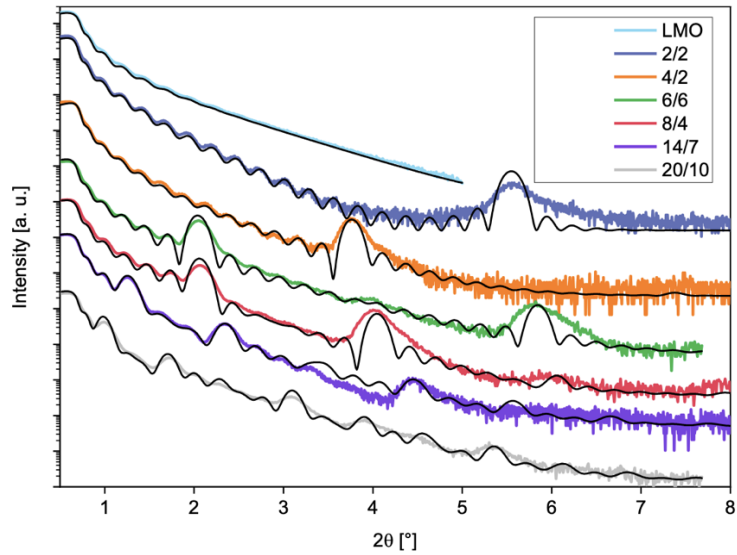

*Figure S3: XRR measurements of the m/n- superlattice films [LaMnO$_3$]$_m$/[SrMnO$_3$]$_n$. The black curves show the best fit for each sample using the algorithms provided by the GenX software.*

| Sample | $d_{LaMnO}$ [Å] | $\sigma_{LaMnO}$ [Å] | | |
|---|---|---|---|---|
| LaMnO | 316(3) | 5.08(3) | | |
| m/n | $d_{Bilayer}$ [Å] | $\sigma_{LaMnO}$ [Å] | $\sigma_{SrMnO}$ [Å] | |
| 2/2 | 16(1) | 6 (1) | 3(2) | |
| 4/2 | 24(1) | 4(1) | 7(3) | |
| m/n | $d_{LaMnO}$ [Å] | $\sigma_{LaMnO}$ [Å] | $d_{SrMnO}$ [Å] | $\sigma_{SrMnO}$ [Å] |
| 6/6 | 23(1) | 3.1(3) | 23(1) | 3.4(7) |
| 8/4 | 31(2)+ | 3.8(6) | 13(1) | 4.5(2) |
| 14/7 | 54(1) | 3.3(3) | 27(2) | 6.7(9) |
| 20/10 | 75(2) | 3.6(2) | 42(2) | 5.6(7) |

*Table S2: Layer thickness, periodicity, and interface roughness as determined by XRR fits. For the 2/2- and 4/2-films only the bilayer thickness $d_{LMO} + d_{SMO}$ is listed because the ratio $d_{LMO}/d_{SMO}$ is determined by the intensity ratio of the superstructure peaks and due to interface roughness σ and instrument resolution, only the first superstructure reflection could be measured for these films.*

SI-1.4 XPS Measurements

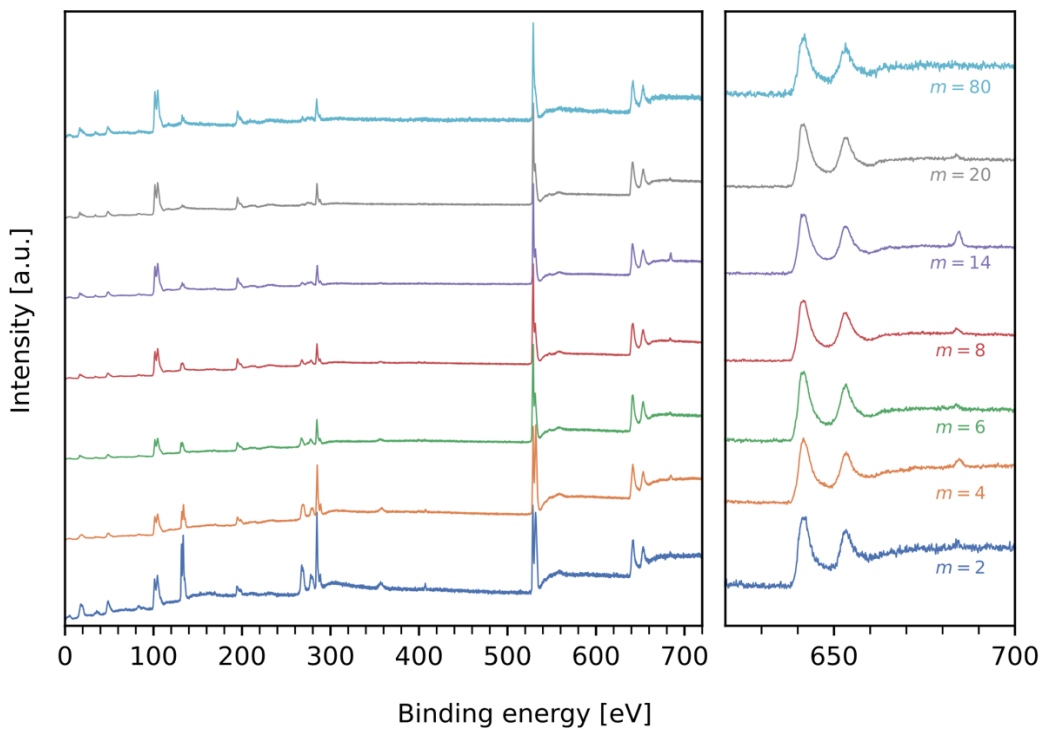

*Figure S4: XPS spectra in a binding energy range of 750 eV – 0 eV. The graph on the right shows an enlargement of the left graph in which the Mn 2p peaks and the F1s peak are more clearly visible.*

X-ray photoelectron spectroscopy (XPS) measurements have been performed on all films using a Kratos Axis Supra spectrometer with a monochromatic Al K$_\alpha$ source. Section from the spectra from all films are shown in Figure S2. The overview spectrum (see Figure S4 (left)) in the binding energy range of 750 eV – 0 eV was obtained with a pass energy of 40 eV and 0.1 eV step size. Higher resolution spectra of the F1s, Mn2p, Mn3p and C1s transitions were acquired with a pass energy of 20 eV and step size 0.1 eV. In addition, an enlarged section of the spectrum is shown on the right, in which the Mn2p and the F1s peak are more clearly visible (see Figure S4 (right)). The binding energy scale has been corrected by shifting the C1s main peak energy to 284.8 eV. Element ratios were calculated by peak intensity ratios after Shirly-background subtraction and transmission correction using the instrument specific relative sensitivity factors (RSF).

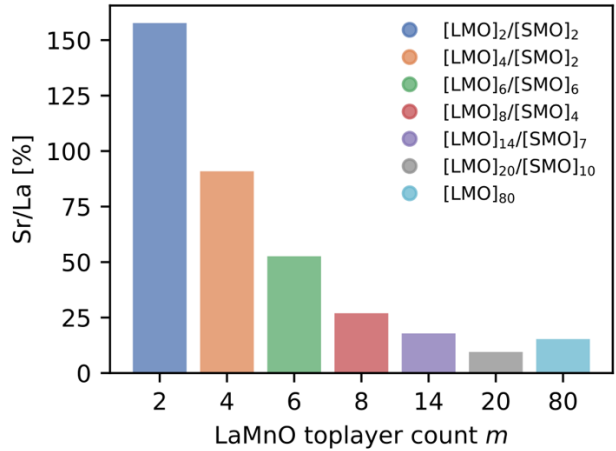

*Figure S5: The Sr/La concentration ratio decreases exponentially with increasing LaMnO top layer thickness due to the exponential decay of information intensity with depth.*

The overview spectra are consistent with a [LaMnO$_3$]$_m$/[SrMnO$_3$]$_n$ layered system. A plot of the Sr/La ratio (Figure S5) shows that the relative strontium signal decreases as expected exponentially with increasing LaMnO top layer thickness due to surface sensitivity of XPS measurements. The inelastic mean free path of the excited e$^-$ is $\lambda = 2.3$ nm in the multilayer system [S2,S3] The Sr/La ratio of 160% for the 2/2-superlattice can be explained by considering the interface roughness of the superlattice. Assuming a local LaMnO top layer thickness of 1 and a SrMnO layer thickness of 3, a Sr/La ratio of 160% is expected according to the lambert beer intensity decay with depth and the calculated IMFP of 2.3 nm.

### SI-1.5 Thermal Resistivity Values

The thermal properties of the films studied here were determined using the results of previous studies [S4,S5], which were obtained from thermal transient reflectometry measurements (TTR) on films prepared in the same system with equal or similar $m$ to $n$ ratios. The $m$ to $n$ ratio can be translated into a unitless interface density $c/\Lambda$ from an effective unit cell thickness $c$ and the superlattice period $\Lambda = m + n$. The estimated values used in this work for thermal resistivity are shown in Figure S7, along with the values from films investigated in the literature [S6]. Further details about the linear model and the two trend lines, depending on the $m/n$-ratio, can be found in [S5].

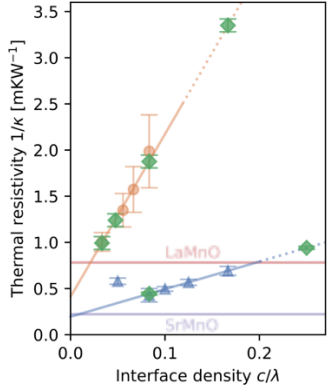

| $c/\Lambda$ | $m$ | $n$ | $1/\kappa$ |
|---|---|---|---|
| 0.25 | 2 | 2 | 0.91(2) |
| 0.166 | 4 | 2 | 3.27(7) |
| 0.083 | 6 | 6 | 0.43(2) |
| 0.083 | 8 | 4 | 1.84(7) |
| 0.048 | 14 | 7 | 1.23(7) |
| 0.033 | 20 | 10 | 0.99(7) |
| — | 80 | 0 | 0.77(7) |

*Figure S6: TTR measurements by D. Meyer et al. [S5] reveal scaling between thermal properties and interface density $c/\lambda = 1/(m+n)$ depending on the $m/n$ ratio. The data, provided by the authors (orange circles and blue triangles), was used to estimate the thermal resistivity for the samples studied in this work (green diamonds).*

*Table S3: Estimates of thermal resistivity $1/\kappa$ for the superlattice samples, based on TTR measurements [S5].*

### SI-1.6 Magnetic Measurements

To examine the stoichiometry and interface charge transfer, the magnetic moment was measured using a Quantum Design MPMS-XL magnetometer (SQUID). Figure S7 depicts the temperature dependent magnetization of the samples. The 32 nm thick LaMnO$_3$ thin film is ferromagnetic with Curie-temperature $T_C = 152$ K which is expected for stoichiometric LaMnO$_3$ due to the epitaxial strain from the SrTiO$_3$: 1.0 at. % Nb substrate [S6].

For the superlattices, the electron transfer between up to 3 unit cells at the [LaMnO$_3$]$_m$/[SrMnO$_3$]$_n$ interface leads to magnetic properties similar to randomly ordered La$_{m/(m+n)}$Sr$_{n/(m+n)}$MnO$_3$, which is the optimally doped La$_{2/3}$Sr$_{1/3}$MnO$_3$ with $T_C \approx 350$ K for the [LaMnO$_3$]$_{2n}$/[SrMnO$_3$]$_n$ superlattices and the antiferromagnetic La$_{0.5}$Sr$_{0.5}$MnO$_3$ for the [LaMnO$_3$]$_n$/[SrMnO$_3$]$_n$ series. [S7] With increasing layer thickness, the magnetic properties gradually change to those of the ferromagnetic bulk LaMnO$_3$ and antiferromagnetic SrMnO$_3$ shown by the decrease in saturation magnetization and increasing coercivity with increasing SrMnO$_3$ thickness in the case of the [LaMnO$_3$]$_{2n}$/[SrMnO$_3$]$_n$ samples (Figure S7 and Figure S8). Therefore, the magnetic properties also support smooth interfaces without intermixing.

Overall, the magnetic properties are in line with those known for MAD grown [LaMnO$_3$]$_{2n}$/[SrMnO$_3$]$_n$ superlattices along with the high-T$_C$ interfacial magnetic phase, giving further justification to using the thermal resistivity values from D. Meyer et al. [S5]

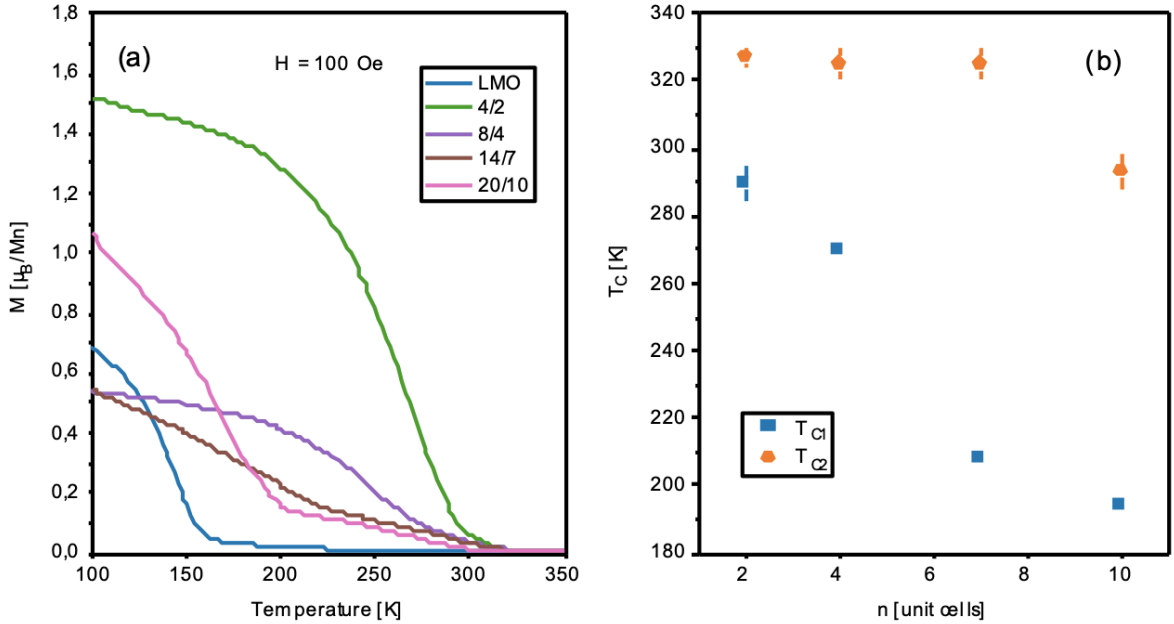

Figure S7: M(T) curves of the [LaMnO$_3$]$_m$/[SrMnO$_3$]$_n$ samples with m = 2n at an external field of 100 Oe (a) and Curie-temperatures T$_{C1}$ of 'bulk' and T$_{C2}$ of the interfacial phase as a function of SrMnO layer thickness.

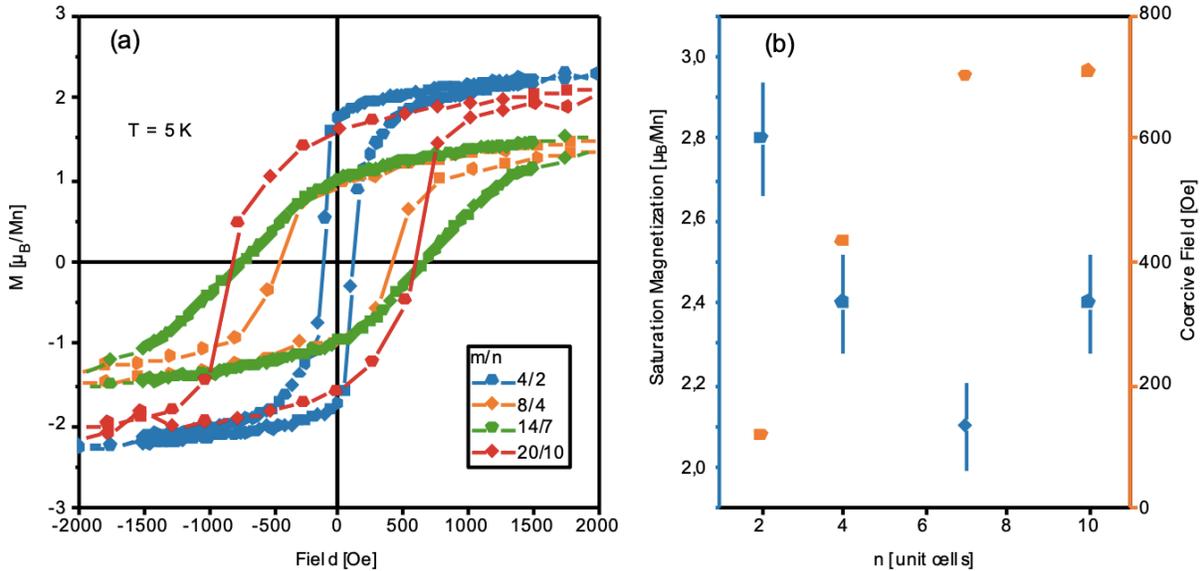

Figure S8: Magnetic hysteresis of the [LaMnO$_3$]$_m$/[SrMnO$_3$]$_n$ with m=2n at T = 5 K (a) and change of saturation magnetization at B$_{ext}$ = 5 T and coactive field with SrMnO thickness n (b).

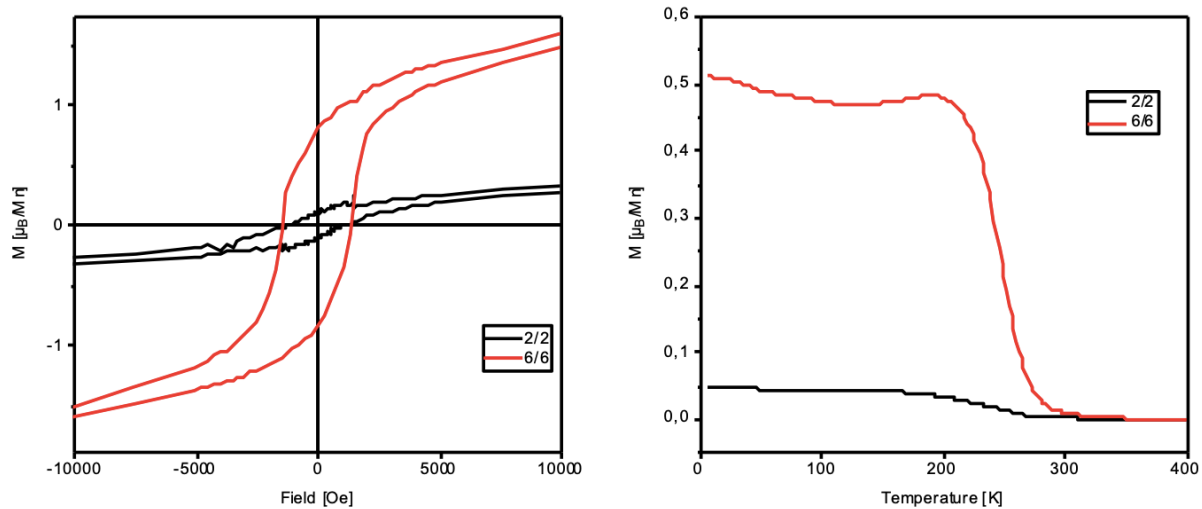

*Figure S9: Magnetic properties of the superlattices with periodicity m=n: hysteresis at T = 5 K (left) and field cooled magnetization at H = 100 Oe (right).*

## SI-2 SEM Cantilever

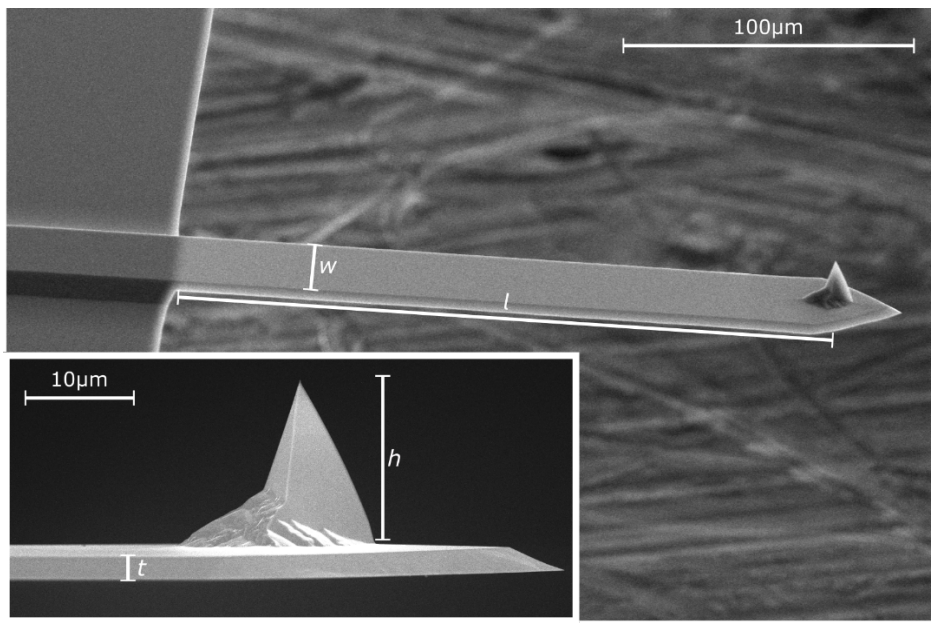

*Figure S10: Scanning electron microscopy images from a cantilever and tip (Nanosensors PPP-LFMR [S8]) used to verify the dimensions stated by the manufacturer.*

## SI-3 Topography and Friction Maps

Figure S11 depicts a representative topography and corresponding friction map for all the seven $[LaMnO_3]_n/[SrMnO_3]_n$ films studied. All maps were measured at room temperature under UHV conditions using an Omicron VT-AFM. The surface roughness $RMS \leq 0.5$ nm is in good agreement with the XRR measurements (see section SI-1.3). We observe low correlations ($r_S <$ 0.15) between the friction force and the height and height gradient maps. All films show local variations in the friction force on a length scale of about 50 nm.

### [LaMnO$_3$]$_2$/[SrMnO$_3$]$_2$

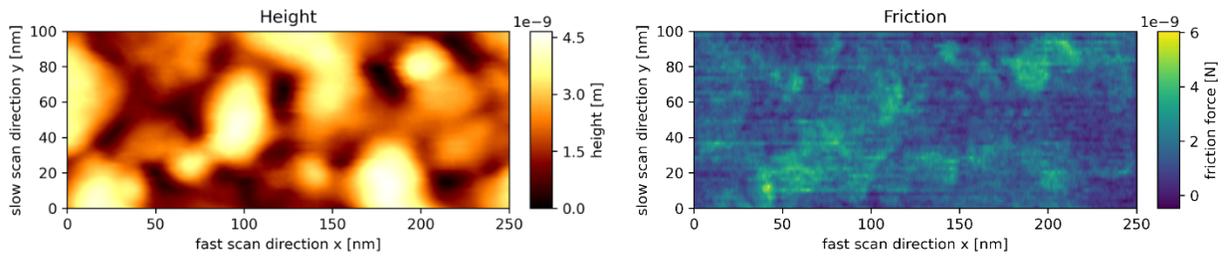

### [LaMnO$_3$]$_4$/[SrMnO$_3$]$_2$

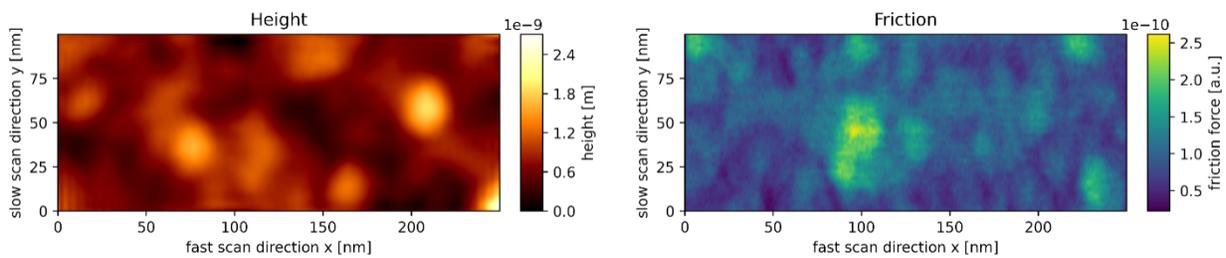

### [LaMnO$_3$]$_6$/[SrMnO$_3$]$_6$

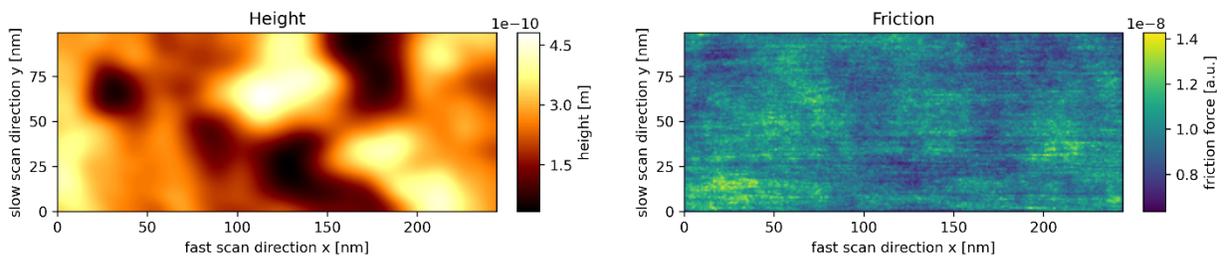

### [LaMnO$_3$]$_8$/[SrMnO$_3$]$_4$

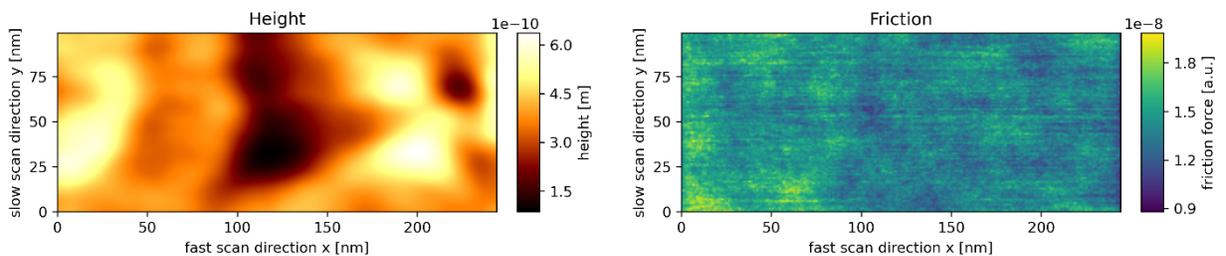

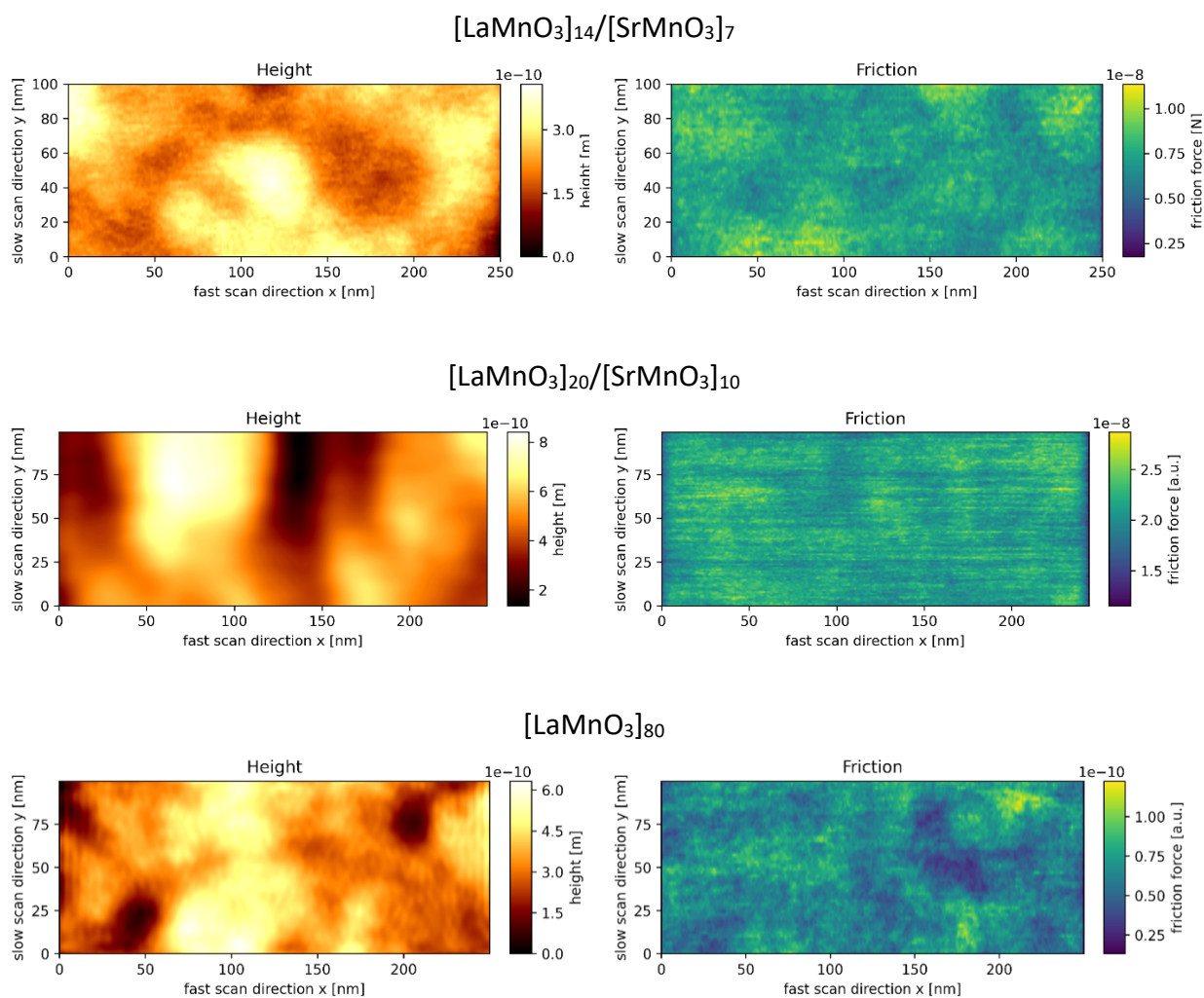

*Figure S11: Topography and friction map, calculated from lateral traces, of the seven superlattice samples studied in this manuscript. All samples, regardless of the m/n ratio, show comparable surfaces with a roughness RMS< 0.5nm, and low correlation coefficients $r_s < 1.5$ between the friction and the height maps.*

## SI-4 Friction versus Normal Force Measurements

### SI-4.1 Friction Force as a Function of Applied Normal Force

The seven plots below show the results of the friction force versus normal force measurements on all seven $[LaMnO_3]_m/[SrMnO_3]_n$ superlattice films. Here, the different point styles indicate different series of measurements on different locations of the sample surface. Each measurement set was performed within a map (100 nm by 250 nm) and the friction forces recorded as the normal force was increased from the initial value and then reduced back to the initial value to check for any changes in tip shape. Independent of the location of the measurement, a linear dependence between the frictional force and the applied normal force can be observed for all films. In contrast, the magnitudes of the measured forces vary strongly from map to map, presumably due to small changes in surface chemistry and/or variations in top layer thickness (see Figure S12). Occasional larger shifts in friction forces ($\pm 10$ nN) are sometimes observed when comparing different measurements made with different cantilevers and are attributed to changes in tip shape.

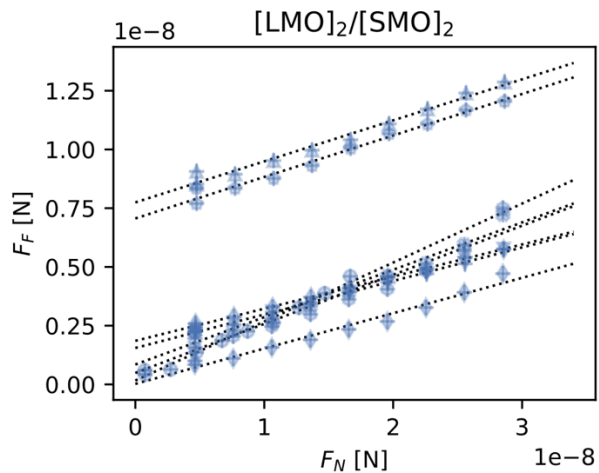
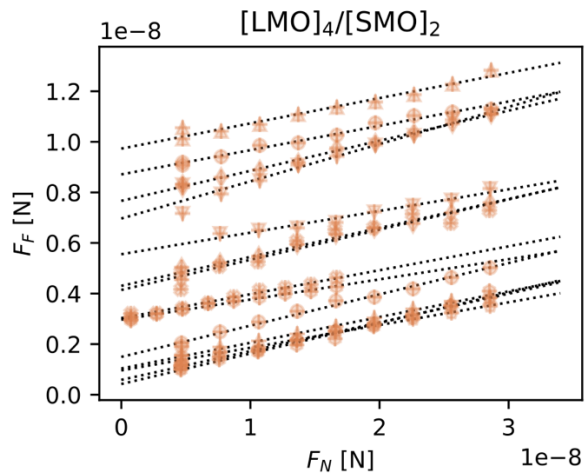
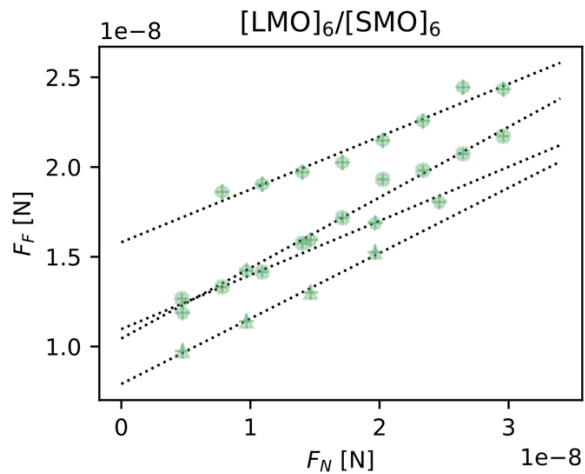
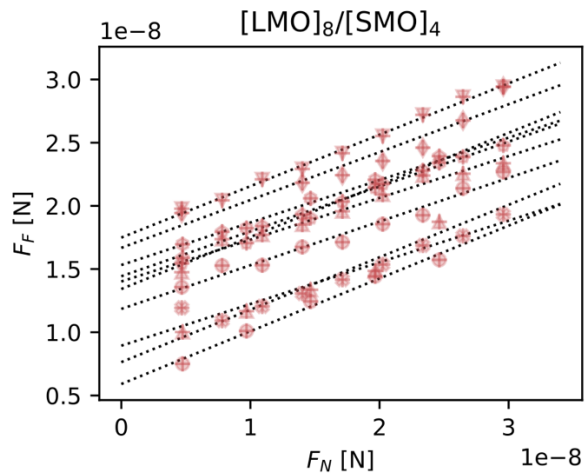
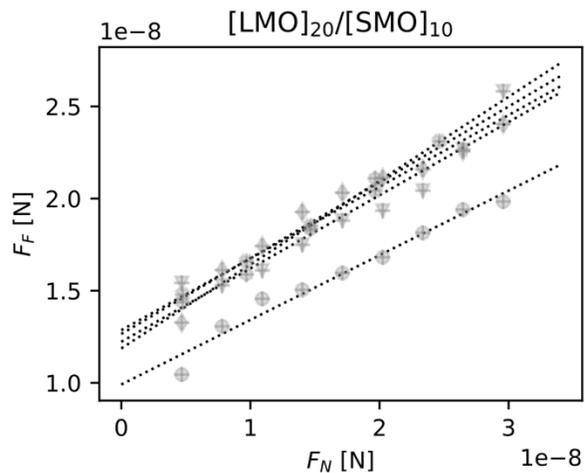
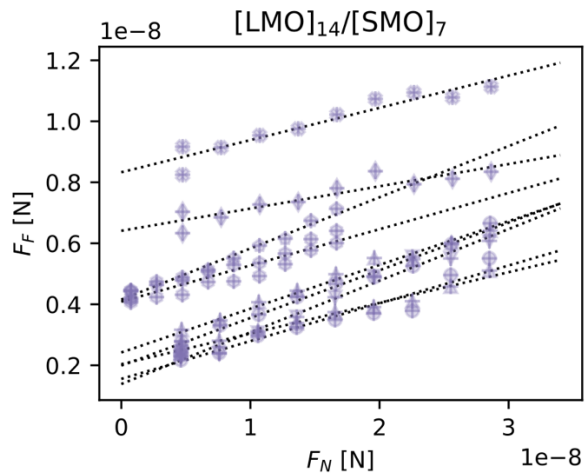

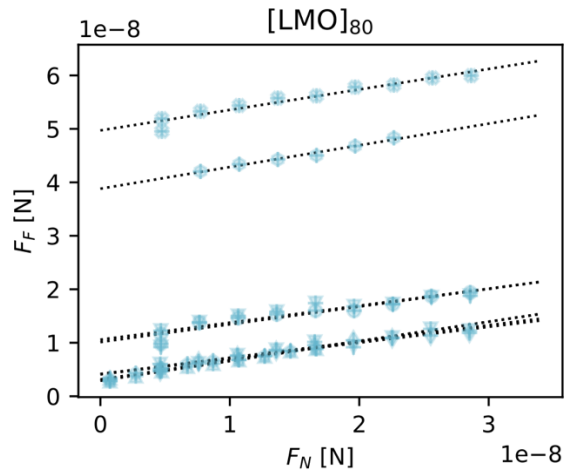

*Figure S12: Friction forces versus applied normal force for all superlattice films. Each dataset corresponds to the average of a single lateral force map (100 nm x 250 nm).*

Linear regressions were performed on each dataset individually, using a modified Amontons relationship [S9]

$$F_F = \mu \cdot F_N + F_{F0}.$$

The friction coefficients (slopes) $\mu$ of the friction force vs applied normal force data are very similar for all of the maps for a given superlattice film, while the y-axis intercept $F_{F0}$ varies strongly from map to map. The friction coefficients of the films not strongly doped with fluorine show a clear dependence increase and saturate with the top LaMnO$_3$ layer thickness (see Figure S13). The two fluorine doped samples $m = 4, 14$ show significantly lower friction coefficients. In contrast, no correlation of the zero intercept $F_{F0}$ with the film thickness or other material properties can be observed (see Figure S13 (right)).

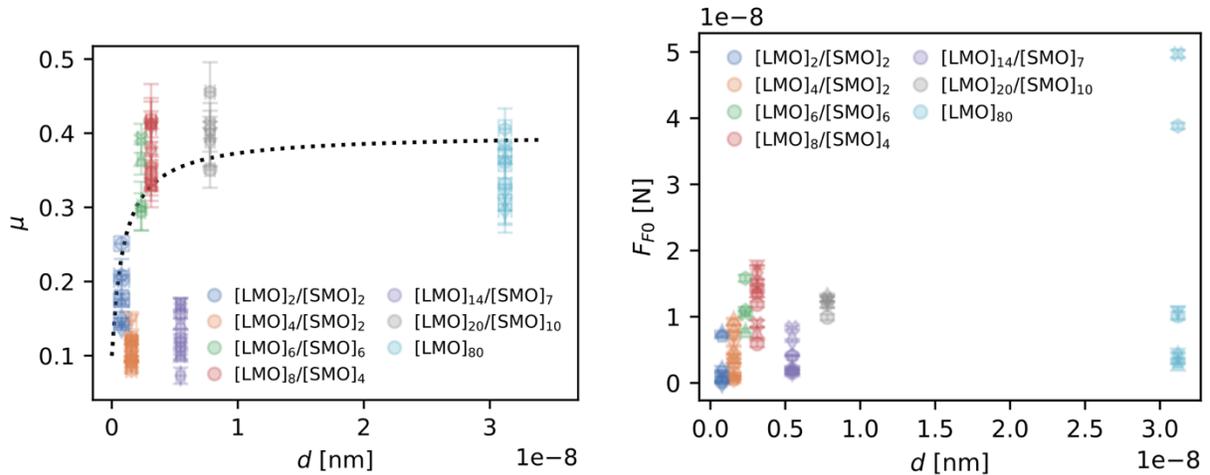

*Figure S13: Friction coefficient μ (left) and friction force at zero applied normal force $F_{F0}$ (right) plotted against the LaMnO top layer thickness for each individual Amontons measurement (see Figure S12). The two fluorine doped samples m=4 and m=14 show much smaller coefficients that we attribute to an altered surface chemistry.*

SI-4.2 Correlation between $F_{F0}/\mu$ and Adhesion

Contact models such as the DMT and JKR model [S10,S11] postulate that the friction force becomes zero when the applied force cancels out the adhesive forces. This means that the value $F_{F0}/\mu$, which is the magnitude of the applied force at which the friction forces extrapolate to zero, should be similar in magnitude to the adhesion force, which can be obtained from the pull-off measurements (force-distance curves). The two forces are compared in the plot below. The good agreement between the two quantities supports the idea that adhesive forces are the cause of the offset in friction force $F_{F0}$.

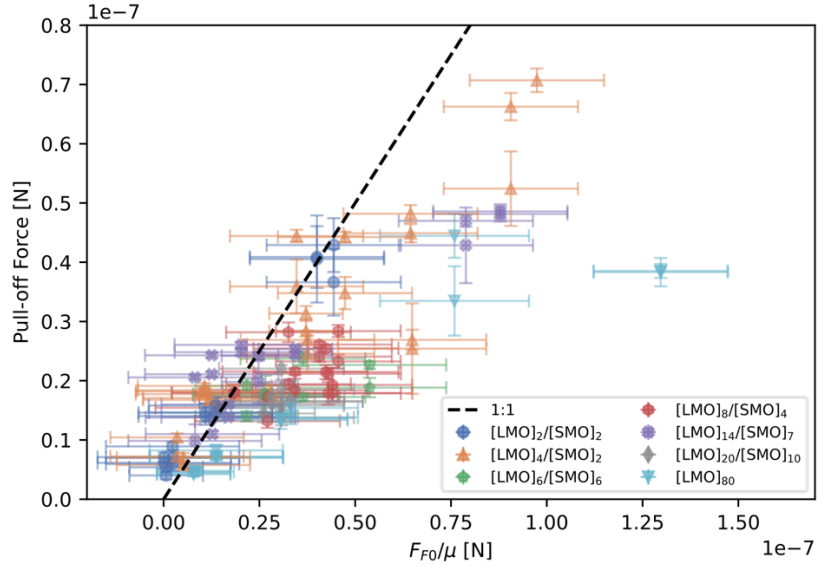

*Figure S14: Pull-off forces obtained from force distance measurements versus the $F_{F0}/\mu$ obtained in the previous section. The dashed line indicates where the two quantities are equal.*

### SI-4.3 Friction Force as a Function of $(F_N + F_{F0}/\mu)$

The good agreement between the pull-off force and the applied force at zero friction suggests that adhesion forces are the cause of the shift in the frictional force $F_{F0}$. Adding this adhesive force $F_{F0}/\mu$ to the applied normal force $F_N$ collapses the friction measurements presented in section SI-4.1 onto a common linear master curve for each superlattice film. The slopes of these curves are referred to as average friction coefficients $\bar{\mu}$ in the manuscript and increases continuously up to a LaMnO top layer thickness of $d = 8 - 10$ nm, after which a saturation to a bulk friction coefficient is observed.

The seven plots (see Figure S16) below show the collapse of the friction data to a single master curve for each superlattice film. Figure S17 depicts the average friction coefficients $\bar{\mu}$ as a function of LaMnO top layer thickness. The friction coefficients initially increase, with increasing LaMnO-thickness and saturate at a film thickness of 5-8 nm to a bulk value.

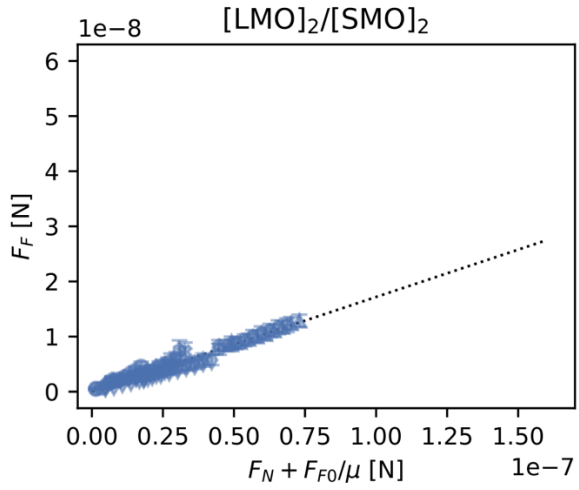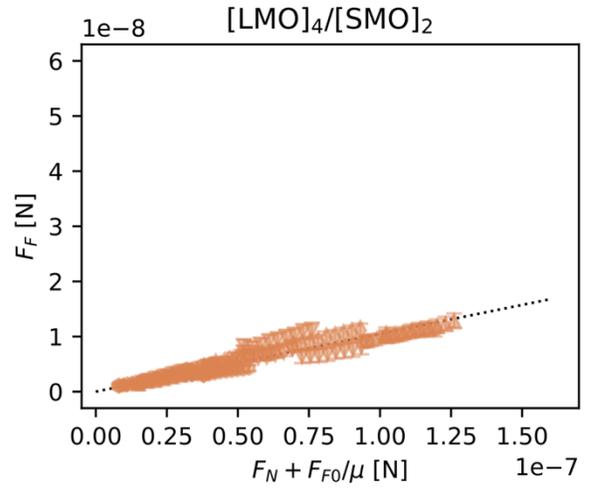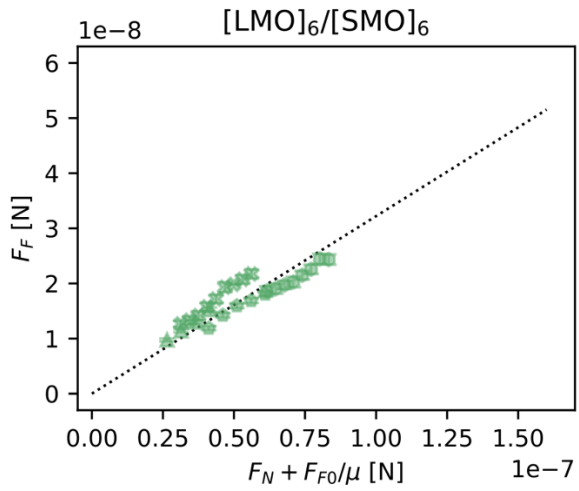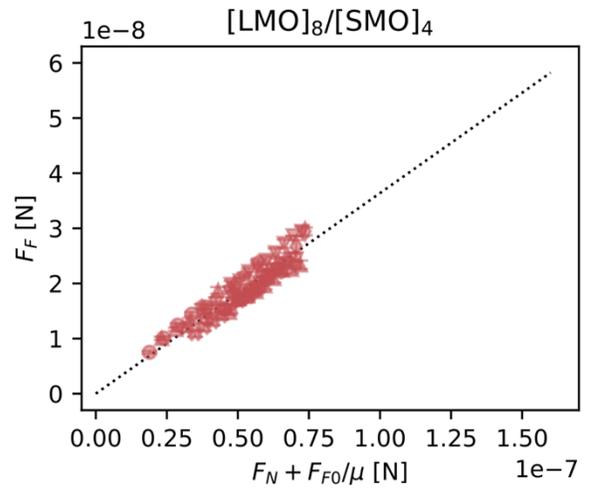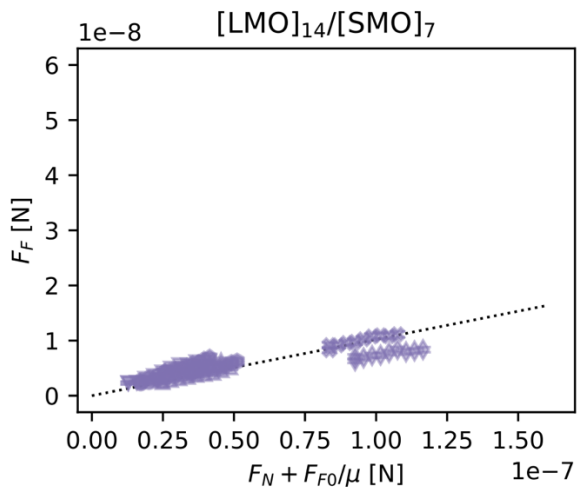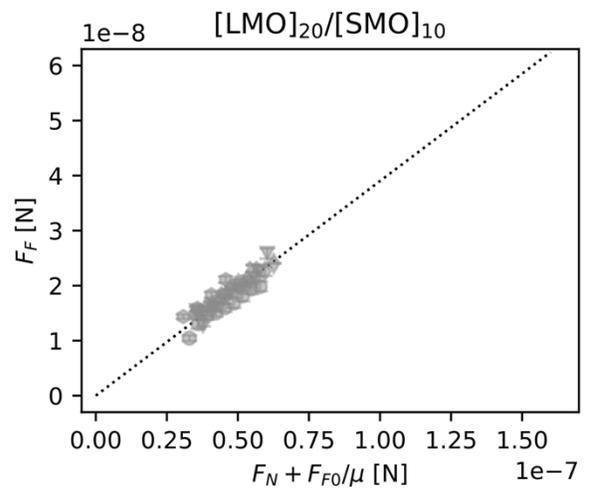

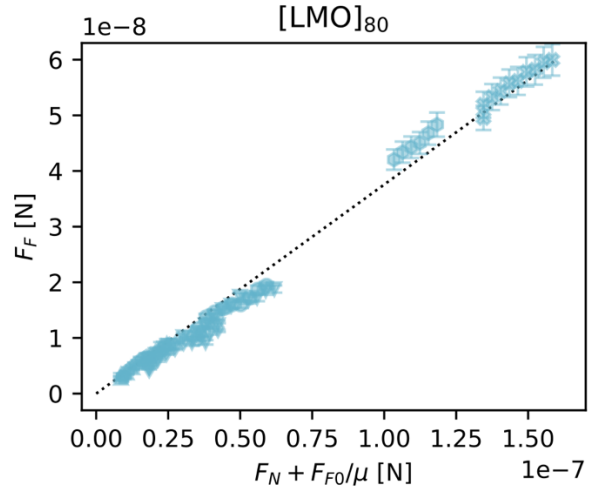

*Figure S15: Friction forces versus $F_N + F_{F0}/\mu$. Each dataset corresponds to the average of a single lateral force map (100 nm x 250 nm). The dashed line shows a linear fit. The slopes obtained from linear fitting $\bar{\mu}$ are depicted in Figure S16.*

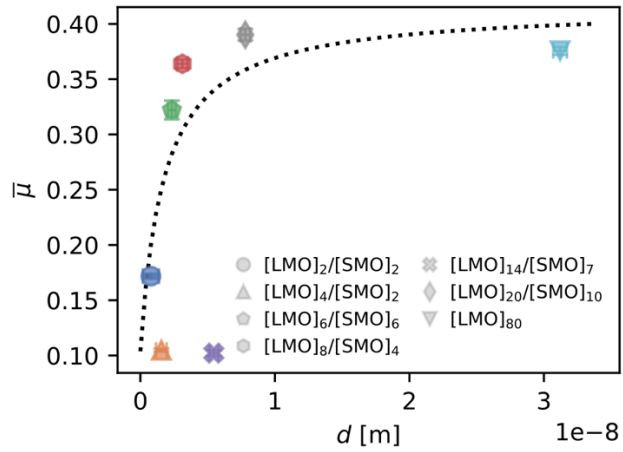

*Figure S16: Slopes $\bar{\mu}$ obtained from linear fitting of the data shown in Figure S15 as a function of LaMnO top layer thickness. Friction coefficients initially increase with increasing top layer thickness and saturate after the film thickness exceeds 5-8 nm. The fluorine dope samples m=4 and m=14 show much smaller friction coefficients compared to sample with similar top layer thickness, that we attribute to an altered surface chemistry.*

## SI-4.4 Master Equation for Friction as a Function of Top Layer Thickness

The functional form of $g(d/a)$ used in Equation 3 of the manuscript was determined by looking for possible scaling between friction force and LaMnO$_3$ top layer thickness. Plotting $(F_F - F_{F0})/F_N$ versus $d/a$ revealed that the measured friction forces collapse onto a single master curve irrespective of the film. This master curve can be reasonably well approximated by a functional form $g(d/a) = d/(d+a)$ an equation that also provides a good fit to the integrated stress fields under a spherical tip [S12,S13] (see Section SI-5). In addition, the functional form is in quantitative agreement with a viscoelastic model developed by Lee et al. [S14], which attributes top layer thickness dependent friction to viscous dissipation inside the evanescent waves set up in the top layer by a vibrating tip.

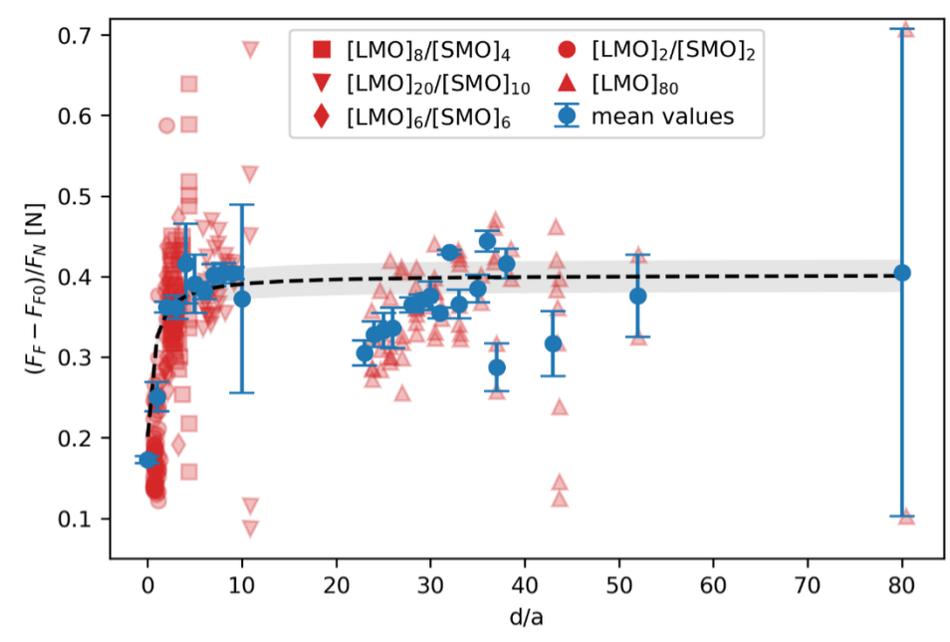

*Figure S17: Plotting $(F_F - F_{F0})/F_N$ versus $d/a$ revealed that the measured friction forces collapse onto a single master curve. This master curve can be reasonably well approximated by a functional form $g(d/a) = d/(d+a)$.*

## SI-5 Elastically strained volume below the tip

An AFM tip in contact with a flat sample surface produces stress fields in the underlying material. Even for nano newton normal loads the stresses under the tip can reach 100 GPa values. However, there is no evidence of plasticity or fracture, so that the material is assumed to respond elastically. Assuming a spherical tip shape the indentation depth $d$ can be estimated by using the Hertz contact model [S13]:

$$d = \frac{a^2}{R} = \left(\frac{9F_N^2}{16E^{*2}R}\right)^{1/3}.$$

The indentation depth $d$, in case of elastic deformation of the surface, is proportional to $a^2$ the interaction area, $R$ the radius of the tip, $E^*$ elastic properties of the materials in contact and the applied normal force $F_N$.

Due to this indentation, a heterogeneous stress field is generated in the underlying material. The exact shape of the stress field depend on the shape of the indenter, the applied normal force, and the elastic properties of the materials. If we again assume a simplified spherical shape of the tip, the stress field under the tip can be calculated using the equation [S12]

$$\sigma_{zr} = 3\,p_m(1+v)\frac{z}{\sqrt{u}}\left[\frac{\sqrt{u}}{a}\tan^{-1}\left(\frac{a}{\sqrt{u}}\right) - 1\right]$$

where $u = \frac{1}{2}[(r^2 + z^2 - a^2) + \sqrt{(r^2 + z^2 - a^2)^2 + 4a^2 z^2}]$, $p_m = \frac{F_N}{\pi a^2}$ the contact pressure and $v$ the Poisson's ratio. The stress fields calculated in this way are long-range and decay radially and in depth only after a few nanometers (see Figure S18).

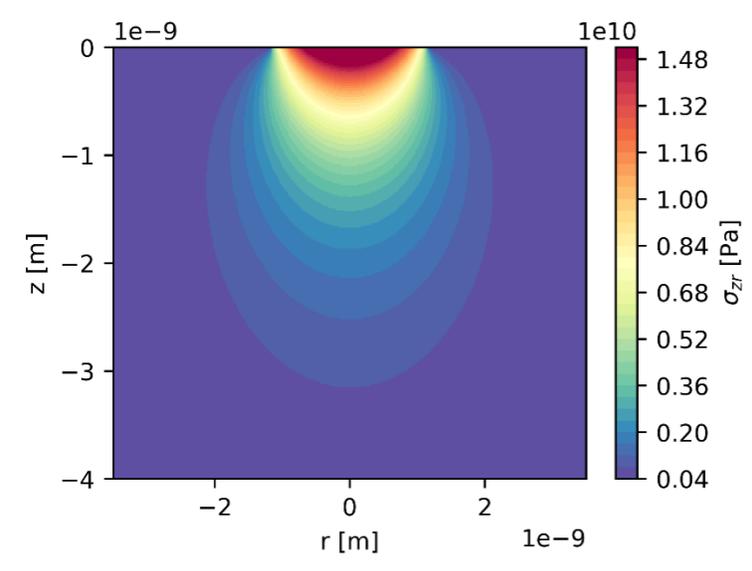

*Figure S18: Stress field generated below a spherical indenter. For the illustration, experimentally reasonable values of $F_N = 20$ nN, $R = 10$ nm, $E^* = 95$ GPa and $v = 0.35$ were used.*

The integrated stress under the tip can be estimated by numerically integrating out to a radius where the stress falls below 10 GPa and then integrating along the z direction (see Fig. S11). The integrated stress under an AFM tip is well described by the function $g(z) = b \cdot (z/(z + a_H))$, where $b$ corresponds to a scaling factor and $a_H$ to an interaction radius that can be estimated from Hertz contact model [S13].

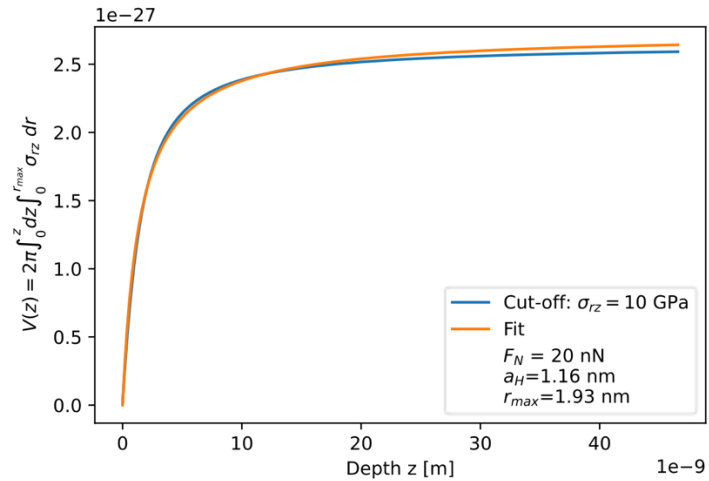

*Figure S19: Integrated stress under an AFM tip. Stresses were integrated out to a distance where the values fall below 10 GPa. The function $g(z) = b \cdot (z/(z + a_H))$ fits the integrated stress reasonably well.*

## SI-6 Bibliography